# EP Aquarii: a new picture of the circumstellar envelope

Pham T. Nhung, Do T. Hoai, Pierre Darriulat, Pham Tuan-Anh, Pham N. Diep, Nguyen B. Ngoc and Tran T. Thai

Department of Astrophysics, Vietnam National Space Center, Vietnam Academy of Science and Technology, 18, Hoang Quoc Viet, Nghia Do, Cau Giay, Ha Noi, Vietnam. *pttnhung@vnsc.org.vn*

**Abstract**  New analyses of earlier ALMA observations of oxygen-rich AGB star EP Aquarii are presented, which complete a previously published analysis and offer a different interpretation of the morpho-kinematics of the circumstellar envelope (CSE). The birth of the equatorial density enhancement (EDE) is shown to occur very close to the star where evidence for rotation has been obtained. Close to the star and where outflows have been observed: their interaction with the gas of the nascent EDE, is seen to play an important role in the development of the wind and the evolution of its radial velocity from 8-10 km s$^{-1}$ on the polar symmetry axis to ∼2 km s$^{-1}$ at the equator. It implies a complex morpho-kinematics: making reliable interpretations with reasonable confidence is difficult. In particular, it questions an earlier interpretation implying the presence of a white dwarf companion orbiting the star at an angular distance of ∼0.4 arcsec from its centre. It proposes instead an interpretation in terms of a standard mass ejection associated with a shock wave leaving a void of emission in its wake. High Doppler velocity wings are seen to consist of two components, the upper velocity end of the global wind, reaching above ±12 km s$^{-1}$, and an effective line broadening, confined within 200 mas from the centre of the star, reaching above ±20 km s$^{-1}$ and interpreted as caused by the pattern of shock waves resulting from the interaction between stellar pulsation and convective cell partition.

**Key words:** stars: AGB and post-AGB – circumstellar matter – stars: individual: EP Aqr – radio lines: stars

# 1 INTRODUCTION

When instruments providing observations of high sensitivity and angular and spectral resolutions became available, mainly the Very Large Telescope (VLT) and the Atacama Large Millimeter/submillimeter Array (ALMA), studies of the circumstellar envelopes (CSE) of Asymptotic Giant Branch (AGB) stars progressed rapidly. Yet, many questions are still unanswered concerning the mechanism that governs the generation and acceleration of the wind, in particular for oxygen-rich low mass-loss rate stars. In contrast with carbon-rich stars, these have a dusty envelope that is more transparent in the visible and near-IR, giving a better view of the innermost dust-forming atmospheric layers (Höfner & Freytag 2019). Their study supports a scenario where the radiation pressure, which accelerates the outflows, is caused by photon scattering on highly transparent iron-free grains, such as corundum ($Al_2O_3$), which form a thin gravitationally bound dust layer close to the stellar photosphere; beyond it, silicate dust condenses on top of corundum cores, speeding up grain growth to the critical size.

Recent high-resolution imaging of nearby AGB stars at visible and infrared wavelengths has revealed complex, non-spherical distributions of gas and dust in the close circumstellar environment, with changes in morphology and grain sizes occurring over the course of weeks or months. At millimetre wavelengths complexity is the main feature that characterizes the morpho-kinematics of the CSE, with important anisotropy and inhomogeneity. Such distribution of atmospheric gas emerges naturally in 3-D hydrodynamical models (Höfner & Freytag 2019) as a consequence of large-scale convective flows below the photosphere and the resulting network of atmospheric shock waves, the dynamical patterns in the gas being imprinted on the dust in the close stellar environment. Global AGB star models are characterized by giant convection cells, which can span over a steradian, and have a lifetime of many years. Together with radial pulsations, with typical periods at a few-month scale, they generate waves of various frequencies and spatial scales, which quickly develop into shock waves giving rise to ballistic gas motions peaking around 2 stellar radii. As the shock waves interact and merge, they produce large-scale regions of enhanced densities in their wakes.

A consequence of the resulting complexity is the difficulty to reconstruct reliably the observed morpho-kinematics of the CSE in both velocity and space, driving its study into proceeding somewhat by trials and errors: successive observations and analyses bring new information that helps refining the picture. EP Aquarii provides such an example with four recent publications by Homan et al. (2018), Tuan-Anh et al. (2019), Hoai et al. (2019) and Homan et al. (2020a). It is an M-type, semi-regular variable star (spectral type M8III), with a pulsation period usually quoted as 55 days but more like 110±20 days (Eggen 1973; Tabur et al. 2009), located at 119±6 pc from the solar system (van Leeuwen 2007; Gaia Collaboration 2018); it shows no technetium in its spectrum (Lebzelter & Hron 1999) and has an effective temperature of ∼3240 K, a mass of ∼1.7 $M_\odot$, a radius of ∼0.77 au (Dumm & Schild 1998) and a mass-loss rate of ∼(1.6±0.4) $10^{-7}$ $M_\odot yr^{-1}$ (Hoai et al. 2019).

The resulting picture that emerges from the four more recent studies of the CSE is a two-component morpho-kinematics producing a Doppler velocity spectrum made of a narrow central peak at star systemic velocity and a broader plateau extending over ∼ ±10-11 km $s^{-1}$. The central peak is associated with an oblate, and the broad plateau with a prolate volume of gas, both expanding radially and sharing a same axis that projects ∼20° west of north and is inclined by only ∼10° with respect to the line of sight. The oblate volume of gas, referred to as Equatorial Density Enhancement (EDE), or occasionally torus or disc, is nearly face on, causing its radial expansion velocity to be poorly defined, ∼2 km $s^{-1}$. It hosts a complex pattern of arcs of enhanced CO(2-1) emission that is interpreted by Homan et al. (2018, 2020a) in terms of a spiral produced by a white dwarf companion, with a mass of 0.65-0.80 $M_\odot$, orbiting EP Aqr at a distance of ∼0.4 arcsec and currently located west of it. The prolate volume of gas, usually referred to as polar or bi-conical outflows, displays maximal velocity along the line of poles but clear density depressions near the poles, the maximum being reached at stellar latitudes between ∼ ±45° and ∼ ±60° while the transition between oblate and prolate components occurs progressively near stellar latitudes of ∼ ±20°.

Evidence for rotation within some 100-200 mas from the centre of the star, with a rotation velocity at km $s^{-1}$ scale, has been obtained from the observation of the emissions of the CO(2-1), $SO_2(16_{6,10}-17_{5,13})$ and $SO_2(4_{2,2}-3_{1,3})$ lines (Homan et al. 2018, 2020a; Tuan-Anh et al. 2019). Observations of the dust induced polarization of the 550-750 nm emission observed at the VLT using SPHERE-ZIMPOL (Homan et al. 2020a) give evidence for a shell of dust surrounding the star and having an inner radius slightly smaller than 100 mas. A study of dust emission in the infrared using the Small Wavelength Spectrometer aboard ISO (Heras & Hony 2005) has given evidence for very low ratios of both gas-to-dust mass and aluminium oxide to silicates relative abundance.

There exists no well-established interpretation of the observed morphology. EDE's are usually described as the result of the gravitational attraction of the stellar wind in the orbital plane of a companion and the polar outflows are often interpreted as episodically produced by the rapid rotation of an accretion disc surrounding this companion and reach very high velocities. Such are the cases of HD 101584 (Olofsson et al. 2019) with highly collimated outflows reaching velocities of $\sim \pm 140$ km s$^{-1}$, of $\pi^1$ Gruis (Doan et al. 2017, 2020; Homan et al. 2020b) with polar outflows consisting of thin hourglass-shaped bubbles reaching a radial expansion velocity of $\sim$60 km s$^{-1}$, of V Hya (Sahai et al. 2022) with high velocity lumpy jets of velocity $\sim$175 km s$^{-1}$, and of R Aqr (Melnikov et al. 2018; Liimets et al. 2018; Bujarrabal et al. 2021) with outflows made of curved central jets with velocity reaching $\sim$240 km s$^{-1}$ surrounded by hourglass bubbles in ballistic expansion with velocity of $\sim$55 km s$^{-1}$. Much closer to the morpho-kinematics of the EP Aqr CSE is that of AGB star RS Cnc, which however displays a clear technetium signal, but is too boreal to be observed by ALMA and has been studied in detail using NOEMA (Winters et al. 2022). It has a two-component morpho-kinematics, quantitatively similar to EP Aqr, with an axis making an angle of $\sim$30° with the line of sight and radial expansion velocities reaching 3-4 km s$^{-1}$ in the EDE and 8-9 km s$^{-1}$ in the bipolar outflows.

Neither EP Aqr nor RS Cnc has been shown to host a companion (but in both cases the presence of an undetected companion cannot be excluded) and no other known AGB star displays a close enough morpho-kinematics to be claimed to serve as a model for their CSE. The question of their genesis remains therefore open and, in the case of EP Aqr, the picture proposed by Homan et al. (2020a), suggesting the presence of an undetected white dwarf companion playing a major role in the formation and structure of the EDE, leaves several questions unanswered that call for further study:

i) High Doppler velocity wings, reaching $\sim \pm 20$ km s$^{-1}$, were interpreted by Tuan-Anh et al. (2019) as two narrow polar structures, referred to as jets, launched from less than 25 au away from the star and building up between $\sim$20 au and $\sim$100 au to maximal velocity. However, subsequent analyses of similar high Doppler velocity wings observed in several other stars, as, for example $o$ Ceti (Hoai et al. 2020; Nhung et al. 2022), favoured interpretations in terms of shocks produced close to the star by pulsations and convective cells, shedding strong doubts on the jet interpretation proposed for EP Aqr. It has therefore become imperative to reconsider the case.

ii) The formation of the bipolar outflows remains unexplained. It may invite interpretations in terms of magnetic fields (García-Segura et al. 2005; Winters et al. 2022), or in terms of interacting winds in a binary (Nordhaus & Blackman 2006; Castellanos-Ramírez et al. 2021), and more specifically in terms of an additional companion orbiting EP Aqr at a distance of ∼10 au (Homan et al. 2020a), but none of these can be solidly justified. The latter authors present a number of relevant suggestive arguments in relation to the observation of a complex pattern of arcs in the emission of the polar outflows.

iii) The formation of the EDE remains equally unexplained. The interpretation proposed by Homan et al. (2020a), in terms of focusing by an assumed white dwarf companion is in principle compelling but raises unanswered questions and is likely to face inconsistencies when details of the observed morpho-kinematics are being considered.

iv) A clear depression of SiO(5-4) line emission is observed in the western blue-shifted octant and interpreted by Homan et al. (2020a) as the result of molecular dissociation produced by the UV radiation of the white dwarf companion. However, to the extent that the companion can be naively expected to sit in the median plane of the disc that it has generated, such dissociation should extend to the red-shifted hemisphere as well and the observed morphology of the depression is difficult to understand in such terms.

The aim of the present article is therefore to shed new light on these issues, by presenting new additional analyses of the emissions of the CO(2-1) and SiO(5-4) lines, in particular in the close neighbourhood of the star. As a brief reminder of the main properties of the CSE of EP Aquarii, we display in Figure 1 intensity maps of the CO(2-1) and SiO(5-4) line emissions, together with Doppler velocity spectra and a schematic description of the axi-symmetric geometry.

## 2  OBSERVATIONS AND DATA REDUCTION

We use archival ALMA observations of the CO(2-1) and SiO(5-4) line emissions of EP Aquarii that were performed in 2016 and 2019 under projects ADS/JAO.ALMA#2016.1.00057.S and ADS/JAO.ALMA#2018.1.00750.S, respectively (PI W. Homan). Observations and data reduction have been described in much detail by Homan et al. (2018, 2020a) and do not need to be repeated here. The latter observations (referred to as TM0) used an extended antenna configuration and were combined with the earlier observations of lower angular resolution by Homan et al. (2020a) with a low weight of 20% by fear of possible artefacts associated with the lack of short spacings. In the present article, in order to better understand their contribution to the combined data, we reduced them separately and report about the results in the present section. The combined data are obtained from Homan et al. (2020a) via the CDS (http://cdsarc.u-strasbg.fr/viz-bin/cat/J/A+A/642/A93). It should be sufficient to summarize their main features in Table 1.

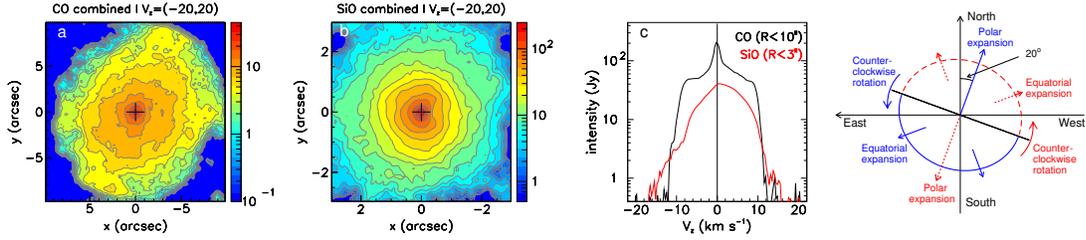

Fig. 1: (a) intensity map of CO(2-1) emission; (b) intensity map of SiO(5-4) emission; (c) Doppler velocity spectra of both CO(2-1) and SiO(5-4) line emissions; (d) schematic of the red-shifts resulting from the tilted axi-symmetric geometry. The colour scales are in units of Jy arcsec$^{-2}$ km s$^{-1}$.

Table 1: Summary of Band 6 observations and data reduction of the CO(2-1) (230.55 GHz) and SiO(5-4) (217.12 GHz) line emissions.

| Reference | Epoch | Line | Beam size (mas$^2$) | Beam PA (°) | Channel width (km s$^{-1}$) | Time on source | Noise rms (mJy beam$^{-1}$) | Maximal Recoverable Scale (") |
|---|---|---|---|---|---|---|---|---|
| Homan et al. (2018) | 2016 | CO(2-1) | 170×150 | −34 | ~0.33 | 70 min (ACA) 230 min (TM1,2) | 1.25 | 15 |
| | | SiO(5-4) | | | | | | |
| Combined data Homan et al. (2020a) | 2016, 2019 | CO(2-1) | 87×79 | +15 | ~0.33 | − | 0.65 | 15 |
| | | SiO(5-4) | 85×79 | +7 | | | | |
| TM0 data present work | 2019 | CO(2-1) | 32×28 | +13 | 0.95 | 99 min | 1.0 | 0.5 |
| | | SiO(5-4) | 32×30 | +22 | 0.34 | | 1.34 | 0.5 |

The 2019 observations were made in two steps, on June 5 and 14, respectively, with a hybrid antenna configuration (C43-10 and C43-9) including ~46 antennas having baseline lengths between 83 and 16000 m. A gap at short spacing (~100-200 m), seen in the baseline length distribution of June 14 (Figure 2a), is due to the absence of four antennas (circled in red in Figure 2b). Four spectral windows were selected to observe line emissions of CO(2-1), SiO(5-4), SO$_2$(4$_{2,2}$-3$_{1,3}$), and $^{13}$CO(2-1). However, the latter two sets of data had insufficient sensitivity to contribute significantly new information and their analysis is omitted from the present work.

The data of June 5 and 14 observations of the CO(2-1) and SiO(5-4) line emissions have been merged with equal weights and grouped in frequency channels corresponding to Doppler velocity intervals of 0.95 and 0.34 km s$^{-1}$, respectively. Natural weighting was used for cleaning, producing maps of 2000×2000 pixels covering ±4 arcsec in both $x$ (pointing east) and $y$ (pointing north), with FWHM beam sizes of 31.7×28.0 mas$^2$ and 32.4×30.0 mas$^2$, respectively. We did not subtract the continuum contribution at visibility level, which would have been improper when looking over the stellar disc because of important absorption. We performed self-calibration but, as it did not result in a significant improvement of the quality of the line images, the data used here do not include self-calibration.

The impact of the lack of short spacing below ∼100 m was found to be different between the CO(2-1) line, with emission extending up to some 10 arcsec angular distance from the star and SiO(5-4) line emission, confined within some 3 arcsec. Moreover, in the case of the CO(2-1) line emission, it was also observed to be stronger near systemic Doppler velocity, where the emission is dominated by a disc extending to large angular distances. The size of the mask, which defines the region in the source plane over which the cleaning procedure is applied, must therefore be chosen accordingly. We studied in detail images obtained with different mask sizes and retained a radius of 2 arcsec for the SiO(5-4) data and of 1 arcsec for the CO(2-1) data, resulting in noise levels of 1.34 mJy beam$^{-1}$ for SiO(5-4) line emission and 1.0 mJy beam$^{-1}$ for CO(2-1) line emission. While the nominal maximal recoverable scale (MRS) corresponding to the C43-10 antenna configuration is 0.46 arcsec, the impact on imaging of the lack of short spacing cannot be simply measured by a single number. As the angular distance from the star increases, it reduces the detected flux first uniformly, then introducing anisotropic distortions of the emission pattern. We studied its effect in much detail, both comparing the produced images with those given by the combined data and by modelling the source as discs of various sizes. While the decline of detected flux starts well below the nominal MRS, important distortions appear only beyond ∼0.6-0.7 arcsec. For the purpose of the present work, we checked carefully that whenever TM0 data were used, proper account was taken of this issue. Channel maps of the SiO(5-4) and CO(2-1) line emissions are shown in Figures S1 and S2 of the supplementary material. Examples of comparison between TM0 and combined data are illustrated in Figures 2, 4, 14 and 15.

We use Cartesian coordinates with the $z$ axis pointing away from Earth in addition to the $x$ and $y$ axes in the plane of the sky. We refer the Doppler velocity $V_z$ to a systemic star velocity of $-33.6$ km s$^{-1}$ in conformity with earlier work, but its value is not known to better than a fraction of a km s$^{-1}$. We define an angular distance from the centre of the star as $R=\sqrt{x^2+y^2}$ and a position angle $\omega$ measured counter-clockwise from north.

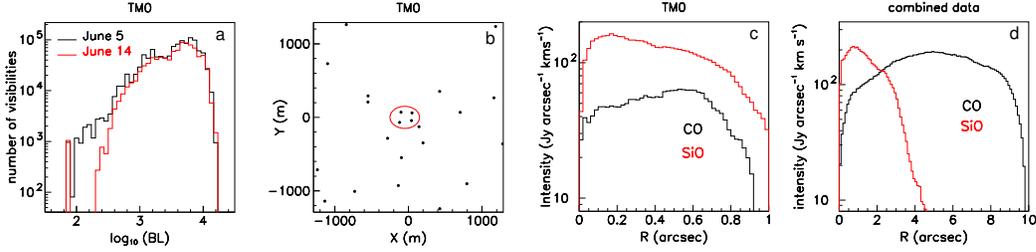

Fig. 2: (a) distribution of the decimal logarithm of the uv distance (in meters) for the different data sets; (b) antenna configuration (centre region); the four antennas circled in red were absent for the June 14 observations; (c,d) radial distributions of CO (black) and SiO (red) emissions using TM0 (c) and combined data (d).

## 3  EQUATORIAL DENSITY ENHANCEMENT

### 3.1  CO(2-1) emission

Evidence for the presence of an equatorial density enhancement (EDE) is known from previous work to rest on the observation of the CO(2-1) and CO(1-0) line emissions exclusively (Hoai et al. 2019). However, a clear distinction between EDE and polar outflows could only be made at distances from the star exceeding ∼2 arcsec. The availability of high angular resolution data makes it now possible to study the EDE morphology at short distances from the star. As illustrated in the upper panels of Figure 3, Doppler velocity spectra integrated over successive rings centred on the star give clear evidence for the presence of the narrow component when $R$ exceeds ∼200 mas. It covers Doppler velocities between 0 and +2 km s$^{-1}$. Remarkably, the neighbouring channels, between $-3$ and 0 km s$^{-1}$ and 2 and 5 km s$^{-1}$, respectively, are depopulated. This may suggest that very close to the star gas is bent toward the equatorial plane, depopulating these channels to the benefit of the narrow component. Channel maps integrated over the corresponding intervals of Doppler velocity are displayed in the lower panels of Figure 3. In the blue-shifted depopulated Doppler velocity intervals, emission is enhanced in the north-eastern quadrant. The intervals of high $|V_z|$ values ($-10$ to $-3$ and 5 to 12 km s$^{-1}$) host outflows expanding along the polar axes; they are discussed in Section 4.

In order to better explore the morpho-kinematics of the nascent EDE, we display in Figure 4 $V_z$ vs $\omega$ maps of the CO(2-1) emission in 100 mas wide rings centred on the star. They show that the EDE is already present in the smaller ring, $R$ between 100 and 200 mas, and its kinematics is dominated by rotation at a few km s$^{-1}$ scale. This evidence for very precocious formation of the EDE had not been made explicitly earlier although Homan et al. (2020a) had remarked that it is suggested by their SPHERE data. In particular, it raises questions in the context of the interpretation given by Homan et al. (2020a), which would imply an inner EDE radius of $\sim$400 mas, corresponding to the orbit of the assumed companion, with strong anisotropy. Unfortunately, with presently available data, it is difficult to tell apart, in the immediate neighbourhood of the star, say below $R\sim$200 mas, the contributions of expansion, of rotation and of phenomena related to the generation of the nascent wind: dust formation and shock waves caused by the interaction between stellar pulsation and convective cell partition (see Section 5).

Farther away from the star, the EDE is seen to subsist up to large distances, as illustrated in the pair of left panels of Figure 5 by the PV maps of the line emission seen in the ring $1.2<R<5.2$ arcsec (see also channel maps of the CO(2-1) emission shown in Appendix in both the $x$ vs $y$ plane, Figure A1, and the $\omega$ vs $R$ plane, Figure A2). These maps, drawn in the $V_z$ vs $R$ and $V_z$ vs $\omega$ planes, respectively, show enhanced emission over all values of $R$ and $\omega$ in the central Doppler velocity interval, $|V_z|<2$ km s$^{-1}$. Similar PV maps drawn in small intervals of $R$ and $\omega$ are shown in Appendix (Figures A3 and A4). In each $(\omega, R)$ bin, with a width of $10°\times100$ mas, we calculate the mean and rms values of $V_z$ over the central Doppler velocity interval, $|V_z|<2$ km s$^{-1}$. Their values are shown as a function of $R$ and $\omega$ in the pair of right panels of Figure 5. They display remarkable uniformity, the rms value staying near 1 km s$^{-1}$ over the whole radial range and the mean value near zero. A small $\omega$ asymmetry of the latter, below $R\sim$3 arcsec, reveals an inclination of the EDE with respect to the plane of the sky consistent with the known inclination of the axis of the CSE with respect to the line of sight: it reaches values of $\sim$0.3 km s$^{-1}$ in the NW direction and of $\sim-0.3$ km s$^{-1}$ in the SE direction. This suggests a radial expansion velocity at the scale of $\sim$0.3/sin10°=1.7 km s$^{-1}$, consistent with the low value of 1.9 km s$^{-1}$ retained in earlier studies (Hoai et al. 2019). The rms deviation of 1 km s$^{-1}$ combines the effect of the known rms flaring angle of $\sim$15°, sin15°$\times$1.7=0.44 km s$^{-1}$ to an intrinsic width of $\sim$0.9 km s$^{-1}$.

### 3.2 SiO(5-4) emission

In order to compare CO(2-1) and SiO(5-4) line emissions in the close neighbourhood of the star, we display in Figure 6 for SiO(5-4) (TM0 data) the same distributions as shown in Figure 3 for CO(2-1). They show a very different picture than those of CO(2-1) emission. The spectra cover a broad range of Doppler velocities, independently from the distance to the star, with no sign of change beyond 200 mas. The narrow component contributes only a very small enhancement. Absorption from the outer layer having reached terminal velocity at 8-10 km s$^{-1}$ is seen on all four spectra. In contrast with CO(2-1) emission, the brightness maps show little structure but reveal the presence of a strong depression of emission at $\sim$400 mas west from the centre of the star. This depression has been abundantly studied and commented upon in the earlier literature and we report on its properties in Section 4, to which we defer possible interpretations. Here we simply remark that the different response to temperature of the two lines can only have small effects. They differ essentially by the value of the energy $E_{up}$ of the upper level, 16.60 K for CO(2-1) and 31.26 K for SiO(5-4). This quantity enters in the expression of the emissivity as a factor e$^{-E_{up}/T}$, where $T$ is the temperature. The ratio between CO(2-1) and SiO(5-4) line emissions of the values taken by this factor for $T$=10 K, 20 K, 50 K and >100 K is 4.3, 2.1, 1.3 and <1.2, respectively.

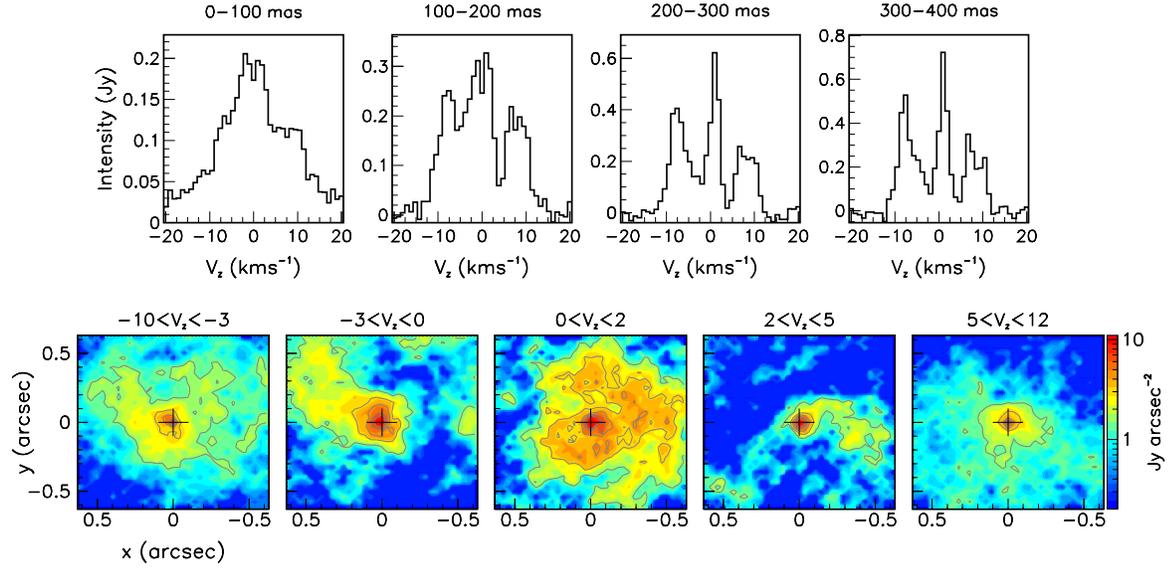

Fig. 3: CO(2-1) line emission, TM0 data. Upper panels: Doppler velocity spectra in successive rings, each 100 mas wide, starting from the circle of radius 100 mas centred on the star. Lower panels: brightness maps averaged over the velocity intervals associated with the structure of the spectra displayed on the upper panels for $R>200$ mas, and indicated on top of each panel.

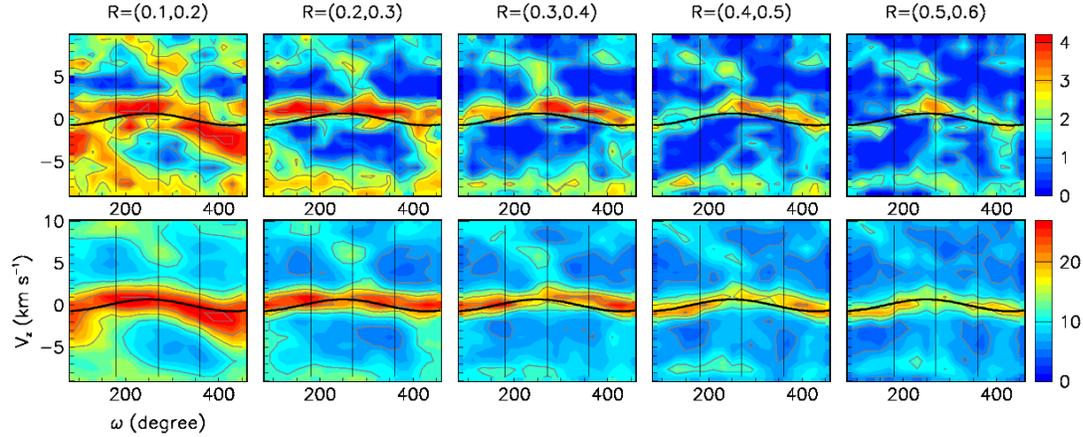

Fig. 4: CO(2-1) emission: TM0 (upper row) and combined (lower row) data. $V_z$ vs $\omega$ maps in 100 mas wide rings as indicated on top of each panel. The vertical lines are at $\omega$ values of 180, 270 and 360°. The sine waves shown as reference correspond to pure rotation with a velocity of 4 km s$^{-1}$. Pure expansion of a same velocity would be phase-shifted by +90°. The colour scales are in units of mJy beam$^{-1}$.

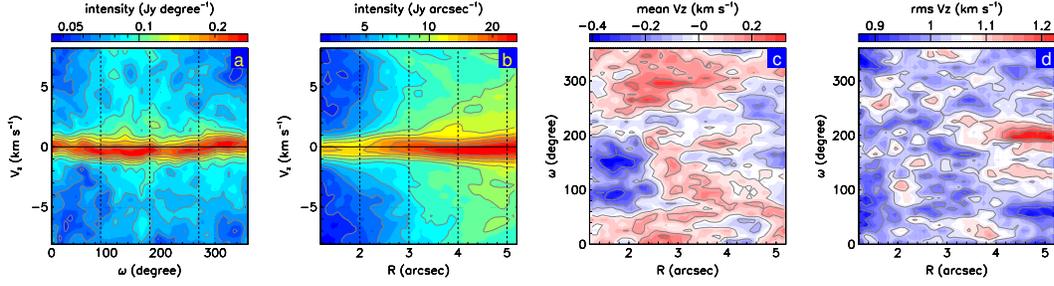

Fig. 5: CO(2-1) line emission, combined data. Left pair of panels: PV maps in the $V_z$ vs $\omega$ (panel a) and $V_z$ vs $R$ (panel b) planes of the projections of the data cube defined as $1.2 < R < 5.2$ arcsec and $|V_z| < 8$ km s$^{-1}$. Right pair of panels: distribution in the $\omega$ vs $R$ plane of the mean (panel c) and rms (panel d) values of $V_z$ over the central Doppler velocity interval, $|V_z| < 2$ km s$^{-1}$.

### 3.3 Patterns of enhanced CO(2-1) emission

The emission of the CO(2-1) line is known from earlier work to display significant fluctuations. Hoai et al. (2019) remarked that both CO(1-0) and CO(2-1) emissions in the EDE display remarkably similar intensity fluctuations, at the level of $\sim \pm 36\%$, dominated by an eastern enhancement in the form of an arc and a southwestern depression, together having the appearance of a spiral. Homan et al. (2018) described similar data as evidence for a bright nearly face-on spiral structure, claiming the first convincing detection of a spiral in the circumstellar environment of an oxygen-rich AGB star, with a higher degree of small scale hydrodynamical structure and wider spiral arms than in carbon-rich environments. In their second paper, Homan et al. (2020a) gave a very detailed description of the EDE intensity fluctuations and concluded that the observed complex emission pattern strongly resembles the typical hydrodynamical gas flow patterns associated with the interaction of a mass-losing AGB star with a companion. They claimed to have identified the flow along the L1 and L3 Lagrange points, before they coalesce into the spiral that manifests at larger length-scales and argued for the presence of a companion located at 0.4 arcsec to the west of the star, probably a white dwarf, with a mass between 0.65 and 0.8 $M_\odot$.

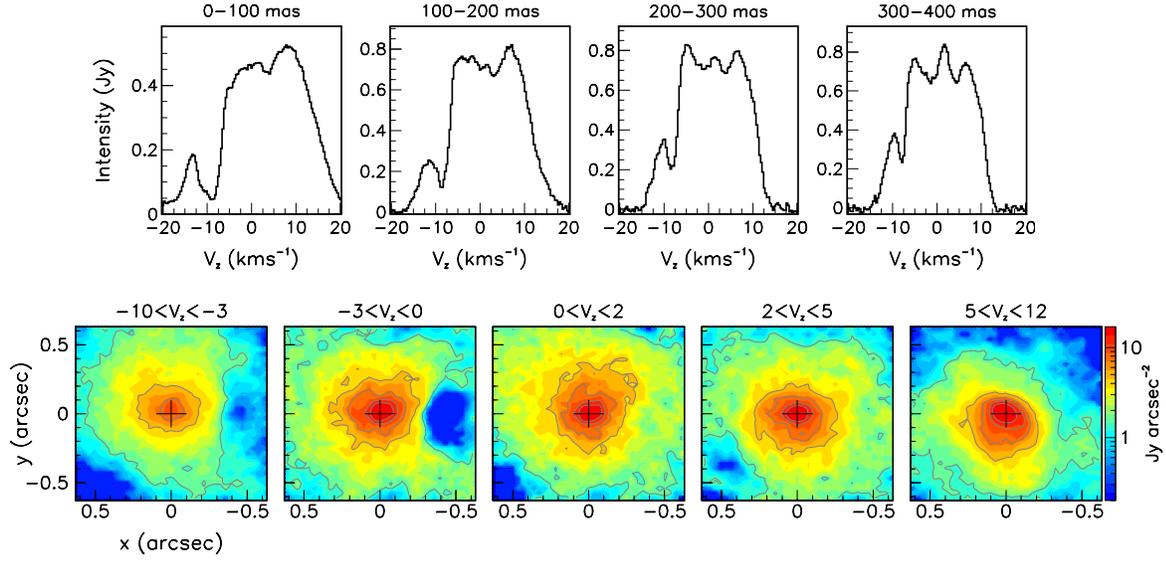

Fig. 6: Same as Figure 3 for the SiO(5-4) emission.

Channel maps of the combined data of CO(2-1) line emission in the twelve frequency channels that bracket the systemic velocity of the star ($-2.0 < V_z < 1.8$ km s$^{-1}$) are shown in Figure 7. They display arcs of enhanced emission. The complexity of the patterns makes a reliable interpretation in terms of well identified individual components difficult. In particular it makes it easy to interpret two partly overlapping arcs as a single one and an arc that is not centred on the star as a spiral arm. The human eye is too subjective a tool for a reliable interpretation of the observed patterns. In this context, the reality of the spiral discovered by Homan et al. (2018) deserves being critically discussed. There is no doubt that identifying such a spiral in the pair of central frequency channels (centred at 0.05 and $-0.27$ km s$^{-1}$, Figure 7) is a natural and reasonable thing to do. However, we note that the neighbouring channels (centred at 0.37 and $-0.59$ km s$^{-1}$, respectively) host patterns suggesting that the source of emission is in a plane inclined with respect to the sky plane: they shine in the NW direction for the red-shifted side and in the SE direction for the blue-shifted side, as expected for a plane perpendicular to the known axis of the CSE. Correcting for this tilt by averaging the brightness over the four central channels, ($-0.75<V_z<0.53$ km s$^{-1}$) we obtain the map shown in the left panel of Figure 8. The identification of a spiral as proposed by Homan et al. (2018) and shown in the central panel of the figure is no longer obvious. Assessing the reliability of the identification of such a spiral is indeed difficult. As illustrated in the right panel of Figure 8, arcs of spirals are excellent approximations to arcs of circles having their centre offset with respect to the centre of the star. But the two halves of a single circle correspond to different spirals winding in opposite directions; a proper spiral winds instead in a single direction and expands indefinitely: an arc having a point of closest approach to the star cannot be described as a single spiral.

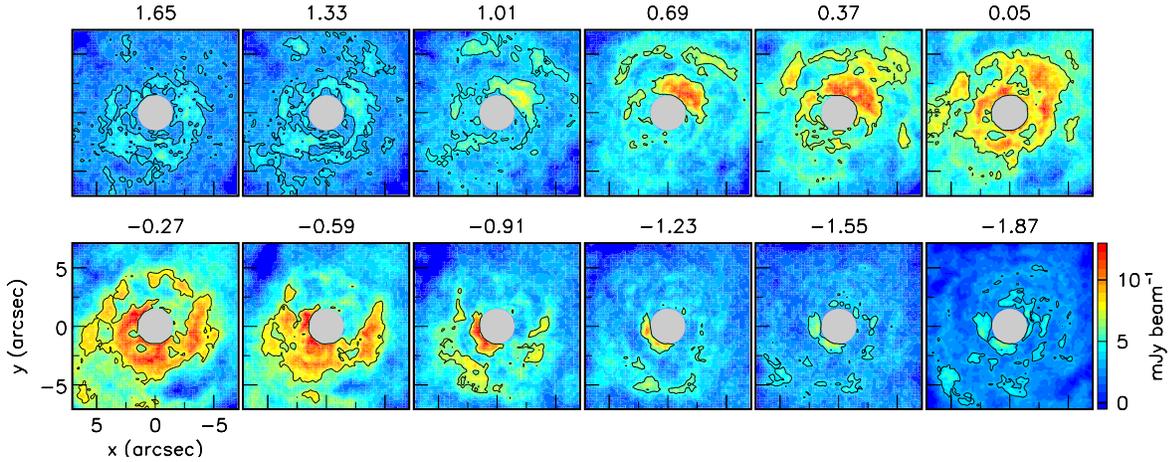

Fig. 7: Channel maps of the CO(2-1) line emission (combined data) over the Doppler velocities (km s$^{-1}$) covering the EDE ($|V_z|<\sim2$ km s$^{-1}$) as indicated on top of each panel. The central region ($R<1.5$ arcsec) is blanked-out for clarity. Contours are shown at 50% of the maximal brightness.

In summary, the EDE is mostly seen in CO line emission as a disc slightly inclined with respect to the plane of the sky, formed within some 100 mas from the centre of the star and having an rms flaring angle of $\sim15°$. Its small inclination prevents a precise evaluation of the morpho-kinematics but evidence has been obtained for rotation in the central region evolving to radial expansion at a low rate of less than 2 km s$^{-1}$ beyond a fraction of an arcsec from the centre of the star. Such a small expansion velocity results probably in part from the low penetration of the stellar radiation into the dense disc, limiting its effectiveness at accelerating dust grains. Assuming a 10 arcsec disc radius (Figure 6 of Homan et al. 2018), it implies that the disc is at least thirty centuries old. Arcs of enhanced emission are observed on the disc but their interpretation as forming a spiral has been disputed.

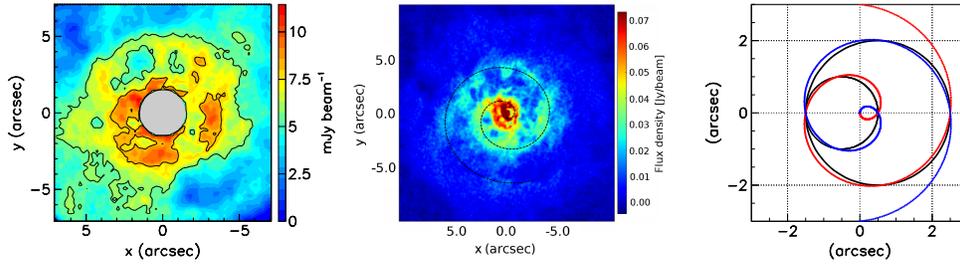

Fig. 8: Left: Brightness of the CO(2-1) line emission (combined data) averaged over the four central panels of Figure 7 ($-0.75 < V_z < 0.53$ km s$^{-1}$). Contours are shown at 50% and 70% of maximal brightness. Centre: from Figure 5 of Homan et al. (2018), central channel of the CO(2-1) line emission. A black dashed Archimedean spiral is over plotted to guide the eye. Right: Comparing two spirals of equation $\rho = 0.5 + \alpha/\pi$ (blue and red) with two circles (black) of radius $R = 1.5$ (2) offset by $-0.5$ ($+0.5$) with respect to the origin. The polar angle $\alpha$ cancels on the axis along which the circles are offset and increases clockwise for the blue spiral and counter-clockwise for the red spiral.

## 4  POLAR OUTFLOWS

### 4.1  CO(2-1) emission

The polar outflows cover a broad region of the sky up to large velocities. This is illustrated in Figure 9, which displays $V_z$ vs $R$ maps integrated over position angles for the blue-shifted and red-shifted hemispheres separately. A simplified schematic of the morpho-kinematics of the CSE is shown in the right panel of Figure 9.

The polar outflows were observed earlier to display density fluctuations at the level of $\sim\pm26\%$, and to reveal both an asymmetry between blue- and red-shifted hemispheres and significant differences between CO(1-0) and CO(2-1) emissions (Hoai et al. 2019). The higher angular resolution of the present data allows for an improved study of the morphology of the enhancements of CO(2-1) line emission. We studied in great detail the data cube (combined data) covering 51 frequency channels (Doppler velocities between $-8.2$ and $+8.2$ km s$^{-1}$) and projected angular distances from the star between 1.2 and 5.2 arcsec (Figures A1 to A4 of the Appendix). In some cases where one might have feared that the steep radial decline of the intensity would have been a source of bias, we examined the brightness divided by a power fit to its radial distribution in the relevant frequency channel, found to have an index between $-0.2$ and $-0.4$. In no case was it considered to be a source of significant bias.

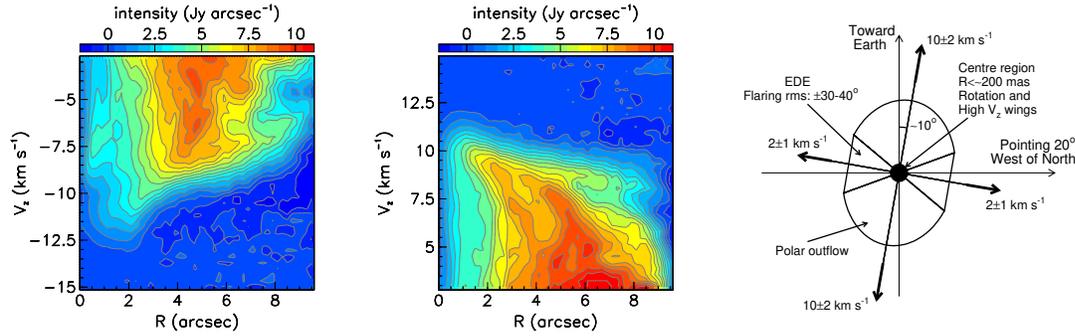

Fig. 9: Left pair of panels: $V_z$ vs $R$ maps of CO(2-1) line emission integrated over position angle for the blue-shifted ($V_z < -3$ km s$^{-1}$) and red-shifted ($V_z > 3$ km s$^{-1}$) hemispheres, separately. Right panel: schematic drawing of the geometry.

Ideally, one would like to describe each channel map as a set of well-identified separated arcs of enhanced emission. We attempted to do so by using an algorithm lumping together the brighter pixels. But we failed to obtain a satisfactory result, the complexity of the patterns preventing reliable pattern recognition. Instead, for each frequency channel, we inspected carefully the different projections of the data cube and tried to describe the observed patterns as simply and reliably as possible. Figure 10 displays a pair of representative examples; they have not been selected to show particularly clear patterns. There are many examples of clearer cases as well as of more obscure cases.

Rather than giving a detailed review of our observations, we simply list the main statements which we can make with confidence: in each map, the pattern of enhanced emission can be described as a set of arcs; they are too numerous, frequently overlapping, to allow for reliable pattern recognition; they can be approximated by arcs of circles covering a few tens of degrees, typically less than 180°; the contrast is small, with a peak-to-valley ratio not exceeding a factor 2; the arcs always wind around the star, implying that the centre of the associated circle is offset by less than a radius, typically half a radius, from the centre of the star; their radii cover a broad range of values, from as small as permitted by the angular resolution to as large as permitted by the size of the maps under study, namely from a fraction of an arcsecond up to over 5 arcsec; very similar patterns are observed in adjacent frequency channels, covering ranges up to a few km s$^{-1}$. These comments complement usefully those made in this context by Homan et al. (2020a).

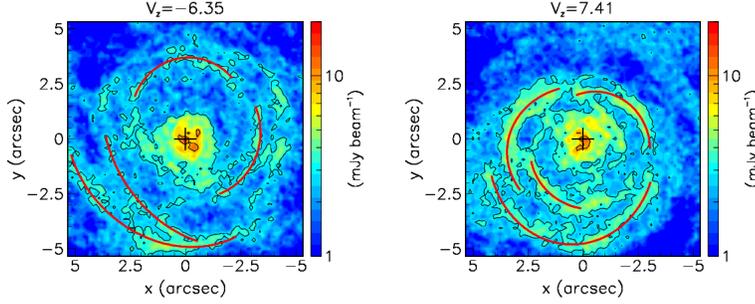

Fig. 10: Examples of patterns of enhanced emission in the blue-shifted (left) and red-shifted (right) hemispheres. Lines guide the eye on some arbitrarily selected patterns.

Finally, we remark that the $V_z$ vs $R$ map shown in Figure 5b (see also Figure A3 of the Appendix) provides precious information on the dependence of the radial expansion velocity on stellar latitude $\alpha$. It was assumed to be of the form $V = V_{eq} + (V_{pole} - V_{eq}) \times \sin^2 \alpha$ in Hoai et al. (2019), where $V_{eq}$ and $V_{pole}$ are the expansion velocities at equator and pole, respectively. As illustrated in Figure 11, a small volume of gas at distance $r$ from the star, at latitude $\alpha$ and having a radial velocity $V$ has coordinates $V_z = V \sin \alpha$ and $R = r \cos \alpha$. Namely, a spherical shell expanding at radial velocity $V_0$ and observed when it reaches a radius $R_0$ is seen as a circle of unit radius in the PV map $V_z/V_0$ vs $R/R_0$. In contrast, a ring expanding at the equator at velocity $V_1$ is seen as a point on the $R$ axis at $V_1/V_0$. Understanding the transition between the polar outflows and the EDE implies understanding the transition between these two regimes. Namely, a shell emitted at some time with velocity $V_0$ at the pole and velocity $V_1$ at the equator is seen as a curve of equation $V/V_0 = V_1/V_0 + (1 - V_1/V_0) f(\vartheta)$ where $\vartheta$ is the polar angle, with $f(0)=0$ and $f(90°)=1$. But $\vartheta$ is equal to the stellar latitude $\alpha$ because the reference radius $R_0$ used to define the axis of abscissa is the distance covered at velocity $V_0$, used as reference velocity on the axis of ordinate. Namely the curve representing the shell in the PV diagram is a direct representation of the function $f(\alpha)$ that describes the dependence on stellar latitude of the expansion velocity. The left panel of Figure 11 shows three examples of function $f(\alpha)$, all three of the form $\sin^n\alpha$, with $n = \frac{1}{2}$, 1 and 2 respectively (in addition to the trivial cases $f(\alpha)$=0 and 1). Comparison with low latitude observations, as shown in the right panels of Figure 11 (see also Figure S3), favours clearly values of $n$ smaller than 1, in contradiction with the arbitrary assumption made in Hoai et al. (2019). Specifically, it suggests that the outflows are close to portions of spherical shells when emitted at large stellar latitudes, and that their radial velocity decreases rapidly at small latitudes where their interaction with the disc becomes important. This is important information for modelling the hydrodynamical flow, however a task beyond the scope of the present work.

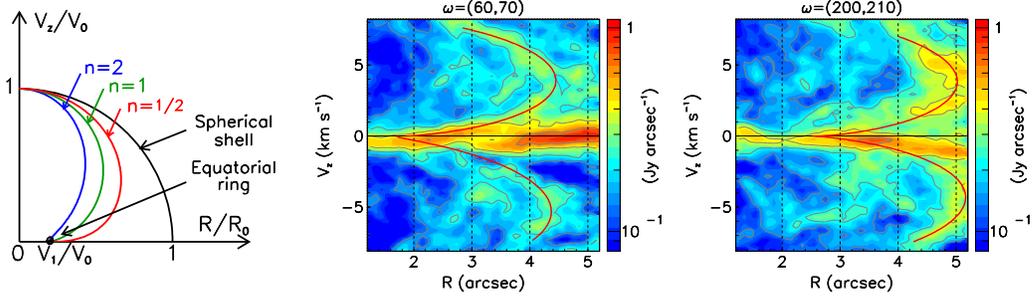

Fig. 11: Latitudinal dependence of the radial expansion velocity. Left: schematics. Centre and right : two examples of $V_z$ vs $R$ maps (CO(2-1) combined data) at $\omega=65°$ and $205°$, respectively (from Figure S3). The red lines in the central and right panels guide the eye on the flows of maximal emission.

Tentatively, the properties of the enhanced patterns of CO line emission in the polar outflows suggest that they take the form of shells of episodic mass ejections and associated shock waves triggered by pulsations of amplitudes enhanced on the surface of convective cells. The confinement of such cells within solid angles at steradian scale limits the patterns observed in the channel maps to finite arcs rather than full circles. Their radial velocity is maximal at the centre of the cell and drops near its edges, causing the intersection of the shell with the sky plane of a given frequency channel to appear eccentric. Moreover, their mean radial velocity increases from the equator to the poles, implying a similar decrease of their density and causing the polar depressions described by Hoai et al. (2019) and the appellation of "bi-conical outflows" adopted by Homan et al. (2020a). Their patterns are best revealed when looking at the $V_z$ vs $R$ distribution of the line emission in a wedge of position angles not exceeding the size of the convective cell (as done in the pair of right panels of Figure 11); integrating over the whole range of position angles or over a large radial distance smoothens out the fluctuations resulting from the episodic nature of the mass ejections, which, as we shall see later, succeed each other at time intervals at the scale of a few decades. At the present stage, such an interpretation is highly speculative but we shall see that further observations described in the reminder of the present section will confirm its likely validity.

## 4.2 SiO(5-4) emission: general features

In contrast to CO(2-1) emission, SiO(5-4) emission is confined within radial distances of ~3 arcsec, as shown in Figure 12, which displays radial distributions for both SiO(5-4) and CO(2-1) emissions. The emissions of the two lines are indeed expected to differ in several respects: the SiO emission decreases with distance to the star, both because of condensation on dust grains and of dissociation by UV interstellar radiation; moreover, it is an efficient shock tracer.

As already remarked, and clearly apparent on the distributions of SiO(5-4) emission displayed in Figure 13, one observes a significant asymmetry between the blue-shifted and red-shifted hemispheres. In particular the former reveals the presence of a strong depression of emission at some 400 to 500 mas west of the centre of the star that is challenging interpretation. In a first step, we limit therefore our considerations to the morphokinematics of the red hemisphere. To illustrate the small inhomogeneity of the data cube, we show in Figure 13 the radial distribution $U(R)$, averaged over position angle and integrated over $V_z$, and the Doppler velocity spectrum $S(V_z)$, of the volume of the data cube defined as $0.5 < R < 2.3$ arcsec, $0 < V_z < 8$ km s$^{-1}$; we then compare the observed intensity to the product $U(R) \times S(V_z)$, properly normalized, in the right panels of the same figure. At the same time as the result reveals clearly the morpho-kinematics of the observed deviations with respect to a smooth distribution, it shows that their amplitude is relatively modest, typically at the level of $\pm 20\%$. Both $U(R)$ and $S(V_z)$ show no significant fluctuation and the lumps of enhanced emissions have typical sizes at the scale of a fraction of an arcsec. In contrast, on the $V_z$ vs $R$ and $V_z$ vs $\omega$ maps, they cover several km s$^{-1}$.

Repeating the same analysis for the blue-shifted hemisphere, we obtain the result illustrated in the lower row of Figure 13. The main difference with the red-shifted hemisphere is a lower intensity, by ~33%, dominated by an important depression in the western quadrant causing a deviation twice as large with respect to the smooth distribution. If it were not for this depression, in the approximation of neglecting fluctuations at the level of $\pm 20\%$, both hemispheres would reveal similar morpho-kinematics, which could be described by a simple nearly spherical model as previously remarked by Homan et al. (2020a). It is therefore important to study it in detail, which we do in the following sub-section.

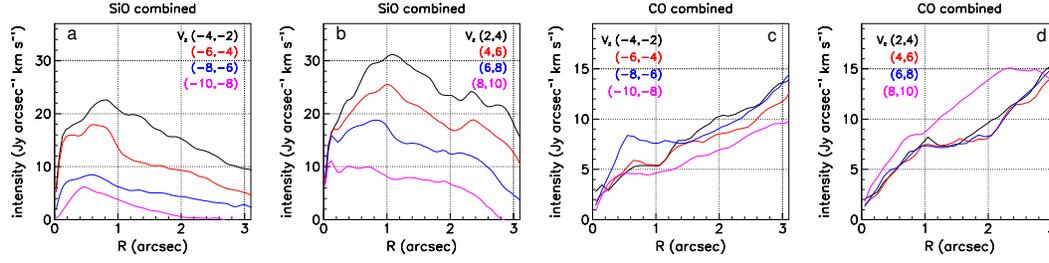

Fig. 12: Radial distributions (combined data), shown separately in the blue-shifted and red-shifted hemispheres, for the SiO(5-4) emission (a and b) and CO(2-1) emission (c and d) integrated over position angle and over Doppler velocity intervals as indicated in the inserts.

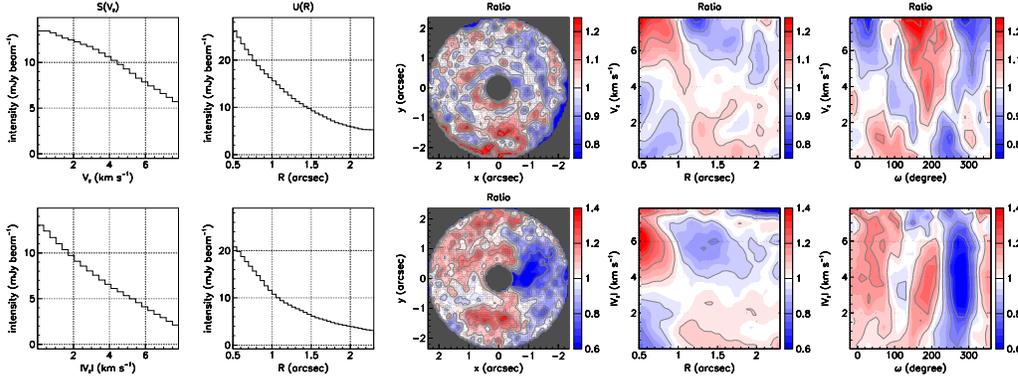

Fig. 13: SiO(5-4) emission (combined data) in the red-shifted (upper row) and blue-shifted (lower row) hemispheres and the radial interval $0.5 < R < 2.3$ arcsec. From left to right: Doppler velocity spectrum, $S(V_z)$; radial distribution, $U(R)$; ratio of the observed intensity to $U(R) \times S(V_z)$, properly normalized, in the $x$ vs $y$, $V_z$ vs $R$ and $V_z$ vs $\omega$ planes, respectively.

Channel maps of the SiO(5-4) line emission are shown as supplementary material (Figure S3). Also shown are PV maps in the $V_z$ vs $R$ plane (Figure S4) and in the $V_z$ vs $\omega$ plane (Figure S5), which display episodic and lumpy emission.

### 4.3 SiO(5-4) emission in the close neighbourhood of the star

Qualitatively, the upper and lower rows of Figure 13 display some similarities: below $R{\sim}1$ arcsec, they both show enhanced emission in the polar outflows, at $V_z{\sim}7$ and $-5$ km s$^{-1}$, respectively, the former covering the southern quadrant and the latter covering a broader interval of position angles, somewhat enhanced in the north-eastern hemisphere. In contrast, beyond $R{\sim}1$ arcsec, emission is enhanced near the EDE.

Figure 14 displays $V_z$ vs $\omega$ maps of SiO(5-4) emission in 100 mas wide rings centred on the star, in the same format as was shown for CO(2-1) emission in Figure 4. Comparing the TM0 and combined data images, and accounting for the different angular resolutions, they clearly reveal the presence in the red-shifted hemisphere of an outflow covering the southern quadrant and of a depression covering the northern quadrant. The picture is less clear in the blue-shifted hemisphere where enhanced emission is seen to cover velocity and position angle values closer to those displayed by the disc, while the western void shows up clearly in both TM0 and combined data images.

While qualitative similarities between both outflows and associated voids are remarkable, as clearly illustrated in Figure 15, important differences must be noted: i) the western blue-shifted depression of SiO(5-4) emission is stronger than the northern red-shifted one, with a peak-to-valley ratio of $\sim$6 compared with $\sim$3; ii) the northern SiO(5-4) depression starts earlier than the western one and is already present in the interval $0.1{<}R{<}0.2$ arcsec where the western depression is not only absent but even partly replaced by a clear enhancement; iii) the southern outflow, covering large Doppler velocities, is emitted in the red-shifted hemisphere close to the line of sight while the blue-shifted outflow, covering small Doppler velocities, is likely to interact with the nascent EDE in the equatorial plane. Indeed, Figure 16 shows that the CO(2-1) emission of the EDE displays a western void of emission clearly associated with the edge of the blue-shifted void of SiO(5-4) emission, as had been previously remarked by Homan et al. (2020a).

These observations underscore the complexity of the dynamics at stake and suggest that the interaction between the blue-shifted outflow and the nascent EDE may play an important role. Particularly puzzling are the properties displayed by the blue-shifted western void of SiO(5-4) emission.

In order to explore these in more detail, we take advantage of its confinement to a relatively narrow interval of position angle, $240°{<}\,\omega\,{<}300°$ to study the distributions of the CO and SiO line emissions in the $V_z$ vs $R$ plane, integrated over $\omega$, which are displayed in the left pair of panels of Figure 17. Both give evidence for a mass ejection in the form of a shell having a mean radius between 0.5 and 1 arcsec and a mean radial velocity between 5 and 10 km s$^{-1}$. The void of emission, for both SiO and CO lines, is clearly seen in its wake, suggesting that it simply results from the radial velocity of the shell significantly exceeding that of the wind that followed it. As a distance of 1 arcsec is covered in the plane of the sky in ~60 years at a velocity of 10 km s$^{-1}$, the mass ejection occurred a few decades ago. We note that the distribution of the CO line emission suggests the presence of an earlier mass ejection that had taken place again a few decades before the more recent one. It is not expected to be visible on the distribution of SiO line emission as the density of SiO gas has significantly decreased at the corresponding radial distance. In line with the interpretation given at the end of Section 4.1 of the patterns observed in the CO line emission of the polar outflows, Figure 17 suggests the occurrence of mass ejections triggered by pulsations of amplitudes enhanced both episodically, with a time scale of a few decades, and in solid angle, at the scale of a steradian corresponding to the presence of a large convective cell. The angular resolution of the present observations cannot resolve the fine time structure associated with the pulsation period, $\sim$110 days.

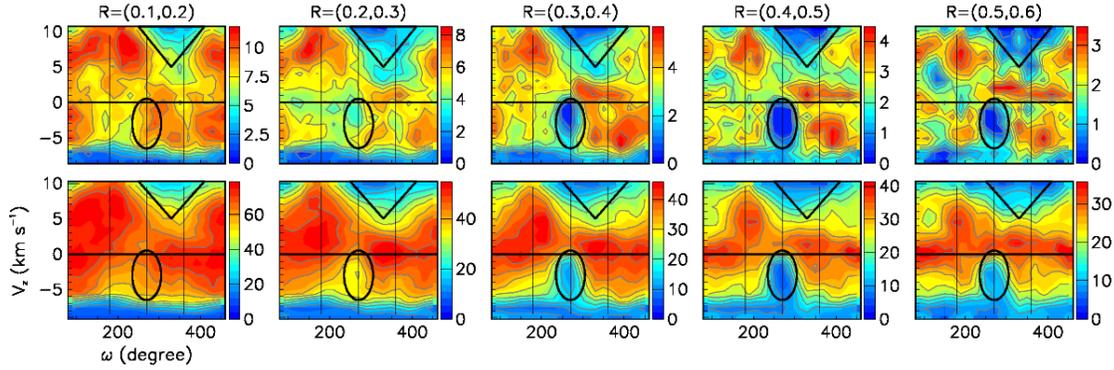

Fig. 14: SiO(5-4) TM0 (upper row) and combined (lower row) data. $V_z$ vs $\omega$ maps in 100 mas wide rings as indicated on top of each panel. Note that the $\omega$ axis is offset by 90°. A triangle and an ellipse, corresponding approximately to the location of the voids of emission, are shown as reference in all panels. The colour scale is in mJy beam$^{-1}$.

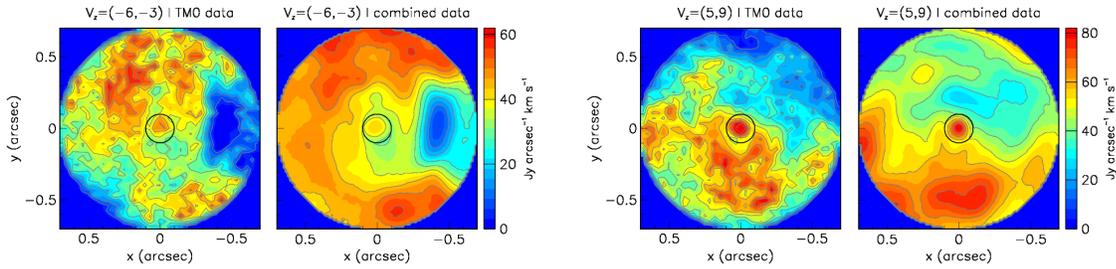

Fig. 15: SiO(5-4) intensity maps integrated over $-6 < V_z < -3$ km s$^{-1}$ for the left pair of panels and over $5 < V_z < 9$ km s$^{-1}$ for the right pair of panels. The intensity is multiplied by $R/0.1$ for $R > 0.1$ arcsec. In each pair, TM0 data are shown left and combined data are shown right.

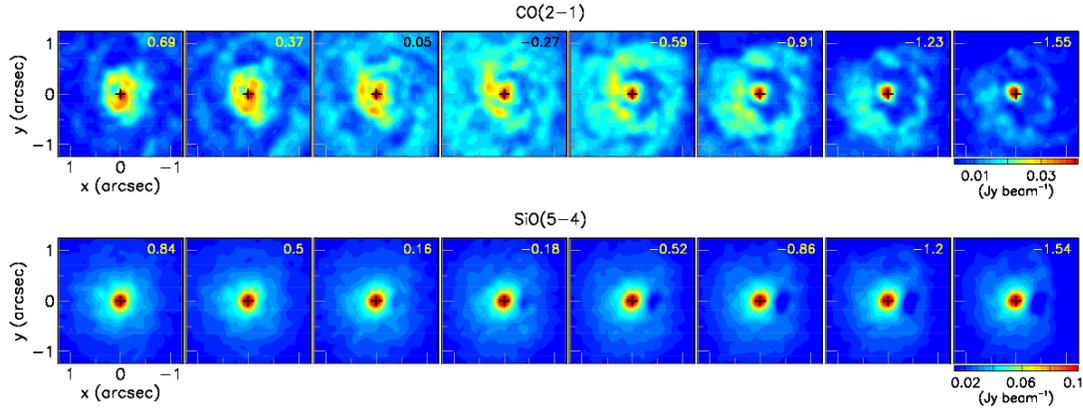

Fig. 16: EDE region, combined data. CO(2-1) (upper row) and SiO(5-4) (lower row) emissions: channel maps centred at the Doppler velocities indicated on top of each panel.

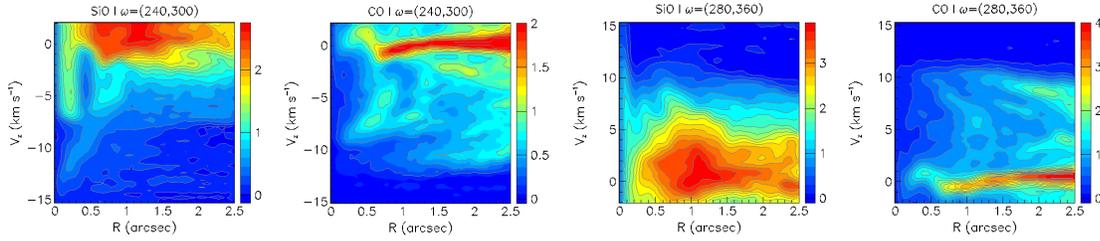

Fig. 17: Distributions of the line emission of SiO(5-4) and CO(2-1) over the $V_z$ vs $R$ plane, integrated over position angles in the interval $240° < \omega < 300°$ (left pair of panels) and $280° < \omega < 360°$ (right pair of panels). In each pair of panels SiO is left and CO is right.

A significant difference between the distributions of SiO and CO line emissions displayed in Figure 17 is the evidence for SiO emission to have resumed after the mass ejection, covering some 100 to 200 mas from the centre of the star, while CO emission has not. Indeed, we saw in Figure 3 that close to the star and in the Doppler velocity interval $-3 < V_z < 0$ km s$^{-1}$ CO line emission is instead enhanced in the north-eastern quadrant. In contrast, we saw from Figure 6 that in the same region of the data-cube, SiO emission covers the whole range of position angles. Understanding the reason for such a difference would require understanding the mechanism that governs the formation of the equatorial disc as well as observations of higher sensitivity and of other line emissions, both of which are lacking. We note that the ejected shell covers latitudes reaching down to the disc, implying that it interacts with it, making it even more difficult to provide a reliable interpretation of the mechanism at stake.

With the aim of exploring the properties of the northern depression, we repeated the analysis made for the western void in the interval of position angles $280° < \omega < 360°$. The result is illustrated in the right pair of panels of Figure 17. Here again, in the range of large Doppler velocities, $V_z > 7$ km s$^{-1}$, the depression is seen in the wake of a shell-like enhancement of emission, much clearer on the CO distribution than it is on the SiO distribution, and an earlier shell is clearly visible on the CO distribution. Also, as for the western void, SiO line emission is seen to have resumed close to the star after the mass ejection, but CO line emission has not. The features displayed by the northern depression are qualitatively similar, but much less clear, than they are by the western void.

While appealing, the interpretation of the western void proposed in the present section leaves a number of questions unanswered, in particular concerning the mechanism responsible for the formation of the equatorial disc. The complexity of the observed features prevents claiming its validity and unicity with strong confidence. But it questions the validity of the interpretation proposed by Homan et al. (2020a) in terms of photo-dissociation by a white dwarf companion: such a companion should have $V_z$ close to zero, like the disc that it is meant to have produced, and should irradiate both the blue-shifted and red-shifted hemispheres. Additional observations of high sensitivity and angular resolution, also including other line emissions, are required in order to ascertain the validity of either of the two scenarios.

Additional figures of relevance to the description of the region of the western depression are displayed in Figures S7 to S9 of the supplementary material.

### 4.4 SiO(5-4) emission: patterns of enhanced emission

In order to compare CO(2-1) and SiO(5-4) line emissions, we must limit the radial range and compensate for the different radial declines. In practice, we consider the ring $0.7 < R < 2.0$ arcsec, also excluding the western sextant $240° < \omega < 300°$ in the blue-shifted hemisphere, and divide the observed intensity by a fit of the $R$ distribution to a form $AR^b$ as was done in Section 4.1. Examples of the resulting maps are shown in Figure 18, one in the blue-shifted hemisphere, with $b$=$-0.20$ for CO(2-1) and $-1.52$ for SiO(5-4), and the other in the red-shifted hemisphere, with $b$=$-0.76$ for CO(2-1) and $-1.20$ for SiO(5-4). The normalised brightness have a mean of unity by construction and an rms width that provides a measure of the intensity contrast, larger for CO(2-1) than for SiO(5-4): 0.30 and 0.31 for the former, 0.15 and 0.22 for the latter, in the red-shifted and blue-shifted hemispheres, respectively. The correlation between the CO(2-1) and SiO(5-4) normalised maps is illustrated in Figure 18 by plotting, for each pixel, the intensity of the latter vs that of the former. A fit of the form $F_{norm}$(SiO)=$F_0$+$G_0 \times F_{norm}$(CO) gives a normalised $\chi^2$, with the values of $\sqrt{\chi^2}$ in the examples shown in Figure 18 being 0.14 and 0.21, in the red- and blue-shifted hemispheres, respectively; the value of $G_0$, which measures the strength of the correlation, is 0.18 in both hemispheres. The pair of examples displayed in Figure 18 calls for a number of remarks of general validity: the maps of CO(2-1) and SiO(5-4) line emissions display more similarities in the red-shifted hemisphere than in the blue-shifted hemisphere; small enhancements of emission are seen on both maps, less contrasted on the SiO map than on the CO map; some are common to both maps, others are seen on only one of the two maps; in contrast to the higher $|V_z|$ range illustrated in Figure 10, clear arcs are rarely observed.

## 5 HIGH DOPPLER VELOCITIES

For a detailed study of the neighbourhood of the star and of the nature of the high velocity wings, we use the high resolution TM0 data in the form of Doppler velocity spectra of SiO(5-4) emission, integrated in 20 mas broad annular rings centred on the star with mean radii increasing from 10 mas to 390 mas in steps of 20 mas. On each of these we define values of the full width measured at 10% of maximum, FWTM, as sketched in Figure 19a. This choice of 10% is a compromise between minimizing systematic uncertainties and staying well above noise and its precise value is of no relevance to the validity of the result. The dependence of FWTM on $R$ gives evidence for confinement within some 200 mas (Figure 19b). It is well described by a fit of the form $21.3+19.0\exp(-R/0.10)$ km s$^{-1}$. The first term corresponds to the velocity range covered by the bulk of the CSE, the second term to a region of high velocities just above the photosphere, as observed in other oxygen-rich AGB stars, and commonly interpreted as resulting from shock waves produced by the interaction of stellar pulsation and convective cell partition (see for example Hoai et al. (2020) and

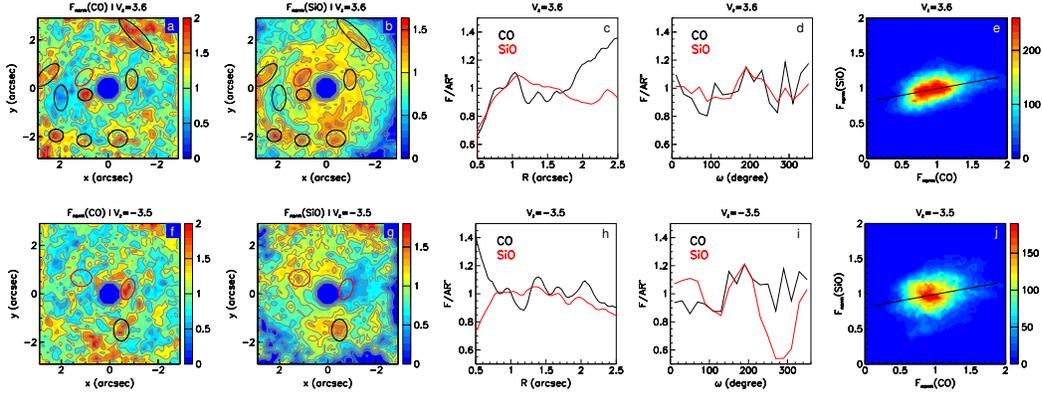

Fig. 18: Comparing CO(2-1) and SiO(5-4) emissions (combined data) in the polar outflows: two examples (in the red-shifted hemisphere, upper row and in the blue-shifted hemisphere, lower row). Left pair of panels: normalized channel maps for $V_z$=3.6 km s$^{-1}$ and $V_z$=−3.5 km s$^{-1}$. We have circled in black some features common to both the CO and SiO maps and in red others that are seen in only one of the two maps. Central panels: normalized $R$ and $\omega$ distributions for CO(2-1) (black) and SiO(5-4) (red), respectively. Right: correlation between the normalised brightness of the two lines; the line is the best fit result (see text).

Nhung et al. (2022)). We also estimate the maximal Doppler velocity $V_{\max}$, reached in the red-shifted hemisphere by the bulk of the expanding CSE, by extrapolating the upper end of the spectrum down to continuum level (essentially zero) using the combined data. The dependence of $V_{max}$ on $R$ and $\omega$ is well described by a fit of the form $V_{\max}$=(1.13−0.11$R$)×[11.1+0.8 sin($\omega$−71°)]. The rms deviation between data and fit is 0.36 km s$^{-1}$. Panels c and d of Figure 19 compare these two factors to the measured mean values of $V_{\max}$, averaged over $\omega$ and $R$, respectively. The dependence on $\omega$ corresponds well to what is expected from the tilt of the CSE axis, pointing ∼20° east of south in the red hemisphere.

This analysis gives therefore evidence for two major contributions to the high velocity wings of the SiO(5-4) Doppler velocity spectrum: one, extending up to some 12 km s$^{-1}$, is simply the upper edge of the velocity distribution of the polar outflows; the other, reaching up to 20 km s$^{-1}$, is confined to the close neighbourhood of the star and corresponds to what is observed in most other oxygen-rich AGB stars where this region of the CSE has been carefully explored. The interpretation given by Tuan-Anh et al. (2019) of these high velocity wings, using data of lower angular resolution, was mixing both contributions. This is clearly illustrated in Figure 20, which displays PV maps of $|V_z|$ vs $R$ for both line emissions and both hemispheres separately. In all four cases there is evidence for a clear separation between the two contributions, even if significant differences exist between the CO(2-1) and SiO(5-4) lines as well as between the red-shifted and blue-shifted hemispheres. Another illustration is given by the intensity maps displayed in the left pair of panels of Figure 21 for $8 < |V_z| < 12$ km s$^{-1}$ (combined data) and in the right pair of panels of the same figure for $12 < |V_z| < 20$ km s$^{-1}$ (TM0 data). The former give evidence for the above-mentioned asymmetry between blue-shifted and red-shifted hemispheres along the projection of the CSE axis, while the latter display instead a small elongation in the perpendicular direction where they are better distinguishable from the bulk. This latter high $V_z$ component seems however different from what we know of other stars by the absence of red-shifted absorption, usually interpreted as evidence for in-falling gas, over the stellar disc (Figure 22). Unfortunately, the lack of reliable light curve measurements prevents commenting on a possible relation with the stellar phase at the time of observations.

## 6 SUMMARY AND DISCUSSION

Collecting the information obtained in both the present work and earlier publications to draw a credible picture of the morpho-kinematics of the CSE of EP Aquarii is challenging. The interpretation given by Homan et al. (2020a), giving a major role to a white dwarf companion currently located at ~400 mas west of the star, has been shown to be not unique and, accordingly, to be less reliable than previously thought, by the results of the analyses made in the present work: the evidence for a spiral enhancement of CO(2-1) emission and the interpretation of the lumps of CO(2-1) emission in the neighbourhood of the star illustrated in Figure A12 of Homan et al. (2020a) have been shown to become invalid when considering a slightly broader Doppler velocity interval. The detailed study of the blue-shifted western depression of SiO(5-4) emission presented in Section 4.3 failed to support an interpretation in terms of dissociation by UV radiation from a white dwarf companion. This does not mean that one can exclude the presence of a companion and its possible role in shaping the CSE. In particular, there is no reason to doubt that EP Aquarii, as most other stars, has several planets orbiting it, some of which may already have been engulfed. But the complexity of the observed morpho-kinematics, and the resulting difficulty to give a reliable interpretation of its many features, prevent claiming a credible model of the mechanisms at stake with sufficient confidence.

In general, the global morpho-kinematics of CO(2-1) emission observed in the present work confirms the results of earlier analyses. The evidence for approximate axi-symmetry about an axis making an angle of $\sim10°$ with the line of sight and projecting $\sim20°$ west of north on the plane of the sky has been strengthened by observations of both the EDE and polar outflows. The latitudinal dependence of the expansion velocity, evolving from $\sim2$ km s$^{-1}$ to $\sim10$ km s$^{-1}$ from equator to poles, has been shown to imply a $\sin^n\alpha$ term with $n$ smaller than unity, in contrast with the assumption of $n=2$ made by Hoai et al. (2019) (Section 4.1). Evidence for enhancements of emission with a typical peak-to-valley ratio below a factor 2, has been found in both EDE and polar outflows, as well as in both CO(2-1) and SiO(5-4) line emissions, often, but not always, in coincidence. Their morphology in the CO line emission of the polar outflows, taking the form of eccentric arcs, suggests that they are caused by shells of mass ejections covering solid angles at steradian scale.

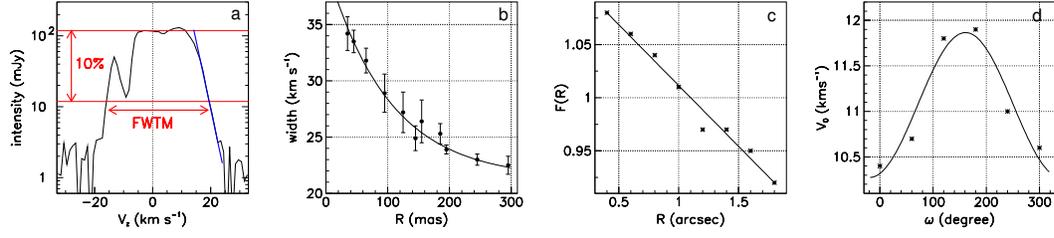

Fig. 19: (a): definition of the quantity FWTM (see text) taking as example the Doppler velocity spectrum of SiO(5-4) line emission measured in the ring $60 < R < 80$ mas (see Figure S6 of the supplementary material). The blue line is used to extrapolate the spectrum (using a linear ordinate scale), its intersection with continuum level defining $V_{max}$ (see text). (b): dependence of FWTM on the ring radius, $R$. The line is a fit $21.3 + 19.0 \exp(-R/0.10)$ km s$^{-1}$. (c): comparison between the measured dependence on $R$ of the mean value of $V_{max}$ and the best fit $F(R) = (1.13 - 0.11R)$. (d): comparison between the measured dependence on $\omega$ of the mean value of $V_{max}$ and the best fit $V_0(\omega) = [11.1 + 0.8 \sin(\omega - 71°)]$.

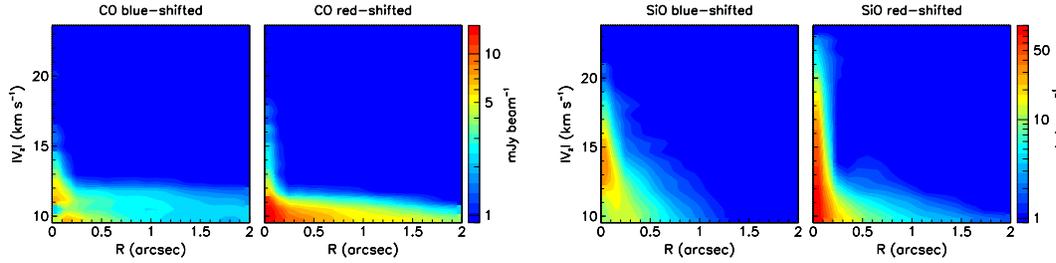

Fig. 20: High Doppler velocities (combined data): PV maps of $|V_z|$ vs $R$ for both hemispheres and both lines separately, as indicated on top of each panel.

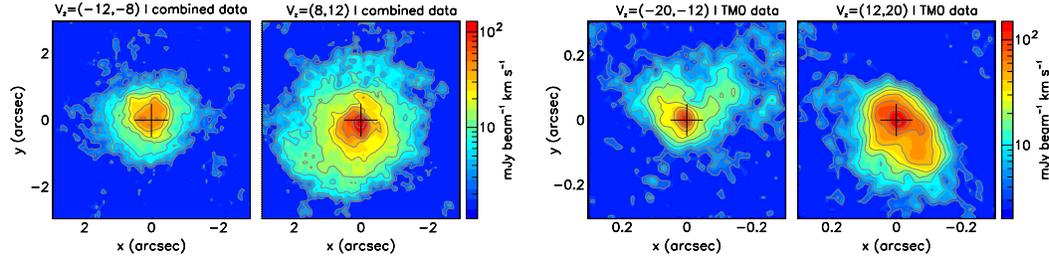

Fig. 21: Left: intensity maps of the SiO(5-4) emission (combined data) measured in the Doppler velocity intervals $8<|V_z|<12$ km s$^{-1}$. Right: intensity maps of the SiO(5-4) emission (TM0 data) measured in the Doppler velocity intervals $12<|V_z|<20$ km s$^{-1}$.

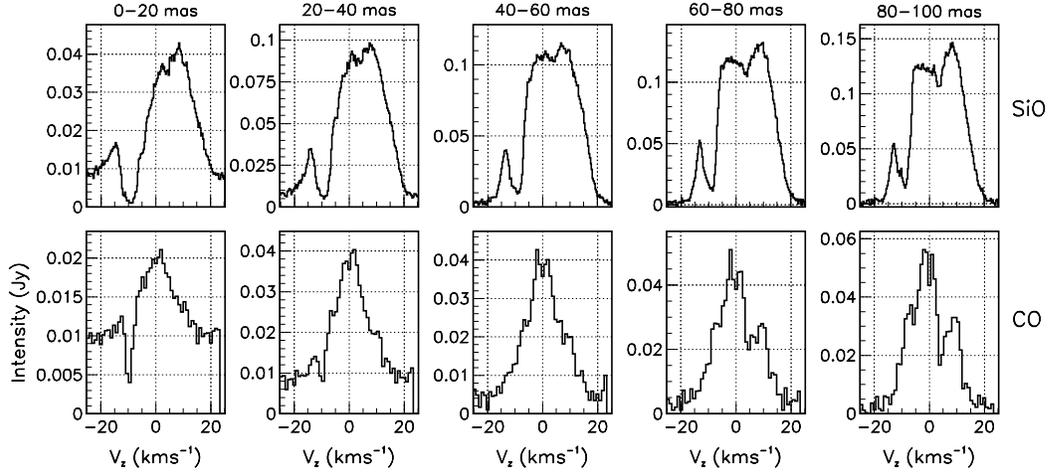

Fig. 22: Doppler velocity spectra of SiO(5-4) (upper panels) and CO(2-1) (lower panels) line emissions (TM0 data) integrated in rings as indicated on top of the upper panels.

The birth of the equatorial enhancement observed in CO(2-1) emission has been shown to occur very close to the star, first dominated by rotation, then by expansion, becoming particularly clear at distances exceeding ~200 mas below which important effective line broadening and dust formation are known to be present (Section 3.1). Lack of evidence for a similar enhancement in SiO(5-4) emission is puzzling; only a very small enhancement is visible at distances exceeding ~300 mas (Section 3.2). This suggests that a disc is formed at short distances from the star probably caused by the presence or the recent engulfment of a companion (Privitera et al. 2016); but this is pure speculation. An intrinsic rms width of nearly 1 km s$^{-1}$ has been found to contribute to the EDE flaring of $\sim$40° FWHM evaluated earlier (Hoai et al. 2019).

The strong difference between the morpho-kinematics displayed by the SiO(5-4) and CO(2-1) emissions (Section 4.2) was understood as being largely due to the different radial ranges being explored, the former declining rapidly as a result of accretion by dust grains and dissociation by the interstellar UV radiation. In a first approximation, SiO(5-4) emission can qualitatively be described by the standard mass-loss mechanism at play in other oxygen-rich AGB stars (Höfner & Freytag 2019), with a typical terminal radial expansion velocity at the scale of 10 km s$^{-1}$ and some prolateness caused by the presence of the EDE.

Indeed, a careful inspection of the morpho-kinematics of the SiO(5-4) line emission in the close neighbourhood of the star has revealed the presence of two distinct broad outflows hosting episodic mass ejections at a time scale of a few decades, covering solid angles at steradian scale, probably associated with shock waves and leaving a depression in their wake: if it were not for the presence of the EDE, the morpho-kinematics of the CSE could be understood as resulting from the standard mechanism of wind generation described by current state-of-the- art hydrodynamical models (Freytag & Höfner 2023). However, the presence of the EDE causes the CSE to take the form of two polar outflows having an increasing radial velocity, and an accordingly decreasing density, when spanning from the equator to the poles.

Such a model, suggested by the results of the present analyses, is at significant variance with that proposed by Homan et al. (2020a). However, in order to claim its superiority, we would need to understand the details of what precisely causes the progressive disappearance of SiO molecules, of what governs the morpho-kinematics of the EDE, of what kind of dust takes part in the acceleration process, of what are the relative roles of stellar pulsation and convective cell partition in giving the initial boost to the wind, of which kind of pattern of shock waves is present.

Only then could one decide with confidence which of the two models is closer to reality. In the present stage, we need additional ALMA observations of high angular resolution and sensitivity to tell them apart. Particularly informative would be observations of the emission of molecular lines that probe preferentially the very close environment of the star, such as from vibrational excited states. New infrared observations of the dust, together with simultaneous monitoring of the light curve, are required to possibly reveal correlation with the stellar phase.

Finally, the presence of high velocity wings, extending up to $\sim$20 km s$^{-1}$, previously noted in earlier work, has been confirmed and shown to include two distinct components (Section 5): upper velocity ends of the polar winds, reaching typically 12 km s$^{-1}$ but with weak tails reaching well above; and effective line broadening confined to distances shorter than 200 mas from the centre of the star. The earlier interpretation given by Tuan-Anh et al. (2019), was failing to make the distinction between these two components. The observed line broadening, consistent with similar observations in all other low mass-loss rate oxygen-rich AGB stars observed with sufficient angular resolution and sensitivity, is believed to be due to lumpy emission of shock waves associated with convective cell partition and stellar pulsation, but a detailed picture is lacking. In general, presently available observations do not allow for telling apart, in the close neighbourhood of the star, the contributions of pulsations, convective cell partition, EDE formation, and dust condensation. EP Aquarii is clearly a star from which one should learn more than currently feasible; it deserves being studied in as many details as possible with state-of-the-art instruments; it still leaves us with many open questions, the answers to which would be of great help to improve our understanding of the mechanisms at stake in the generation and development of the wind of oxygen-rich AGB stars.

**Acknowledgements** We thank Dr Ward Homan and Dr Anita Richards for clarifications concerning their earlier work and for sharing with us their understanding of the morpho-kinematics of EP Aquarii. We also thank Dr. Stefan Uttenthaler for discussion related to the properties of the CSE. We thank the anonymous referee for comments which helped with improving the presentation. This paper makes use of the following ALMA data: ADS/JAO.ALMA#2016.1.00057.S and ADS/JAO.ALMA#2018.1.00750.S. ALMA is a partnership of ESO (representing its member states), NSF (USA) and NINS (Japan), together with NRC (Canada), MOST and ASIAA (Taiwan), and KASI (Republic of Korea), in cooperation with the Republic of Chile. The Joint ALMA Observatory is operated by ESO, AUI/NRAO and NAOJ. We are deeply indebted to the ALMA partnership, whose open access policy means invaluable support and encouragement for Vietnamese astrophysics. The combined data are obtained from Homan et al. (2020a) via the CDS (http://cdsarc.u-strasbg.fr/viz-bin/cat/J/A+A/642/A93). Financial support from the World Laboratory, the Odon Vallet Foundation and the Vietnam National Space Center is gratefully acknowledged. This research is funded by the Vietnam National Foundation for Science and Technology Development (NAFOSTED) under grant number 103.99-2019.368.

**Appendix A:**

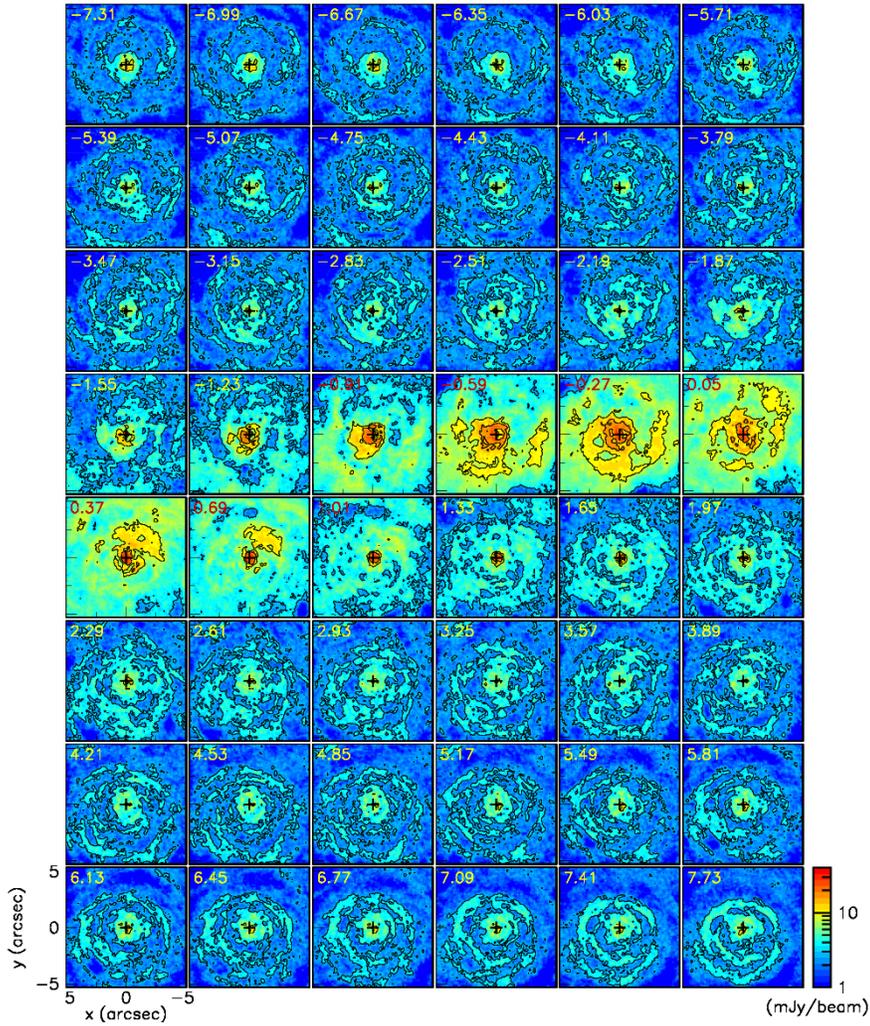

Fig. A1: Channel maps of CO(2-1) emission (combined data). Central Doppler velocities (km s$^{-1}$) are indicated for each channel on the upper-left corner of the panel.

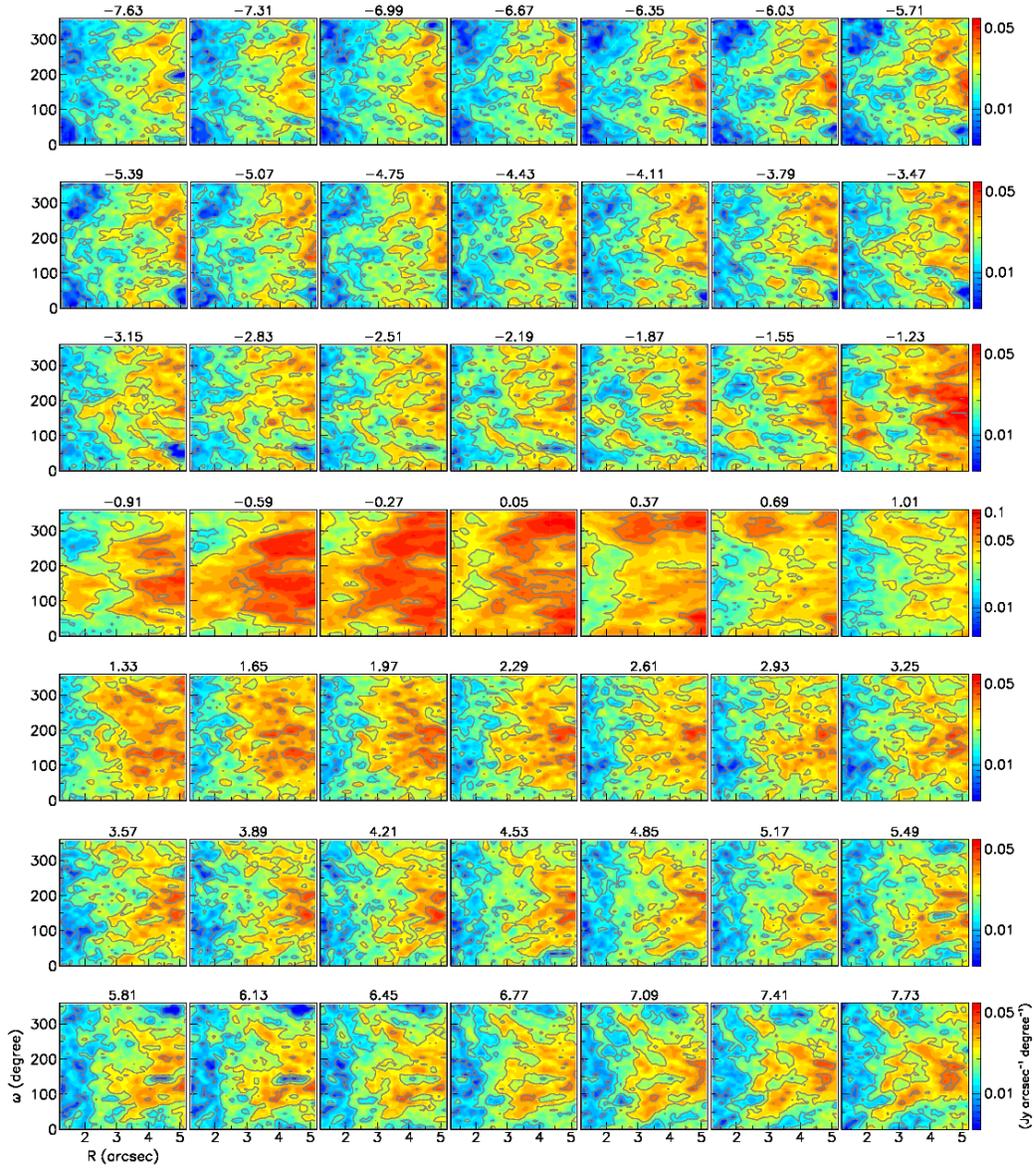

Fig. A2: Channel maps of CO(2-1) emission (combined data) in the $\omega$ vs $R$ plane. Central Doppler velocities (km s$^{-1}$) are indicated for each channel on top of the panel.

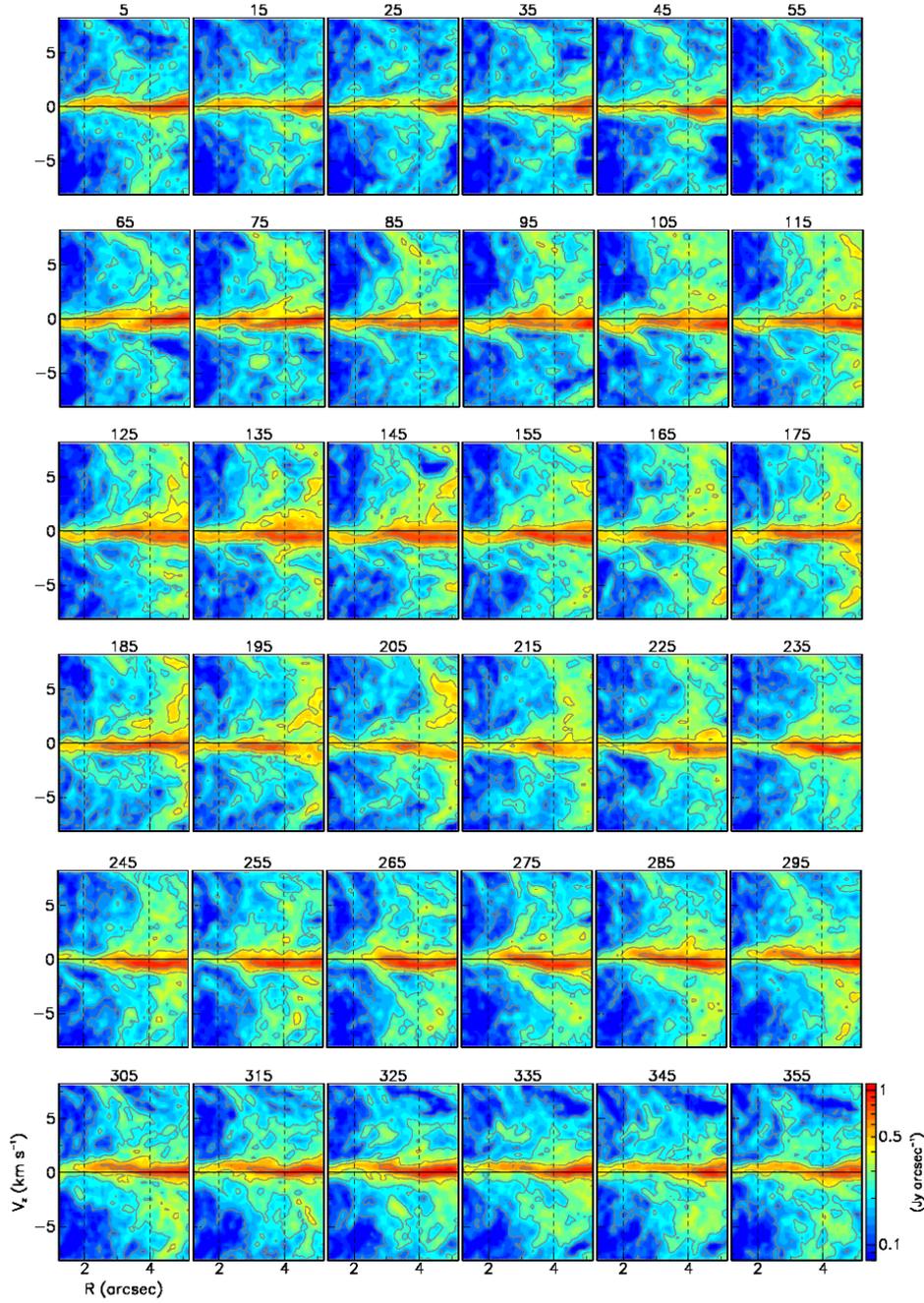

Fig. A3: Projections of the data cube of CO(2-1) emission (combined data) in 36 intervals of $\omega$ (the central value in degree is indicated on top of each panel) on the $V_z$ vs $R$ plane ($|V_z|$<8 km s$^{-1}$, 1.2<$R$<5.2 arcsec).

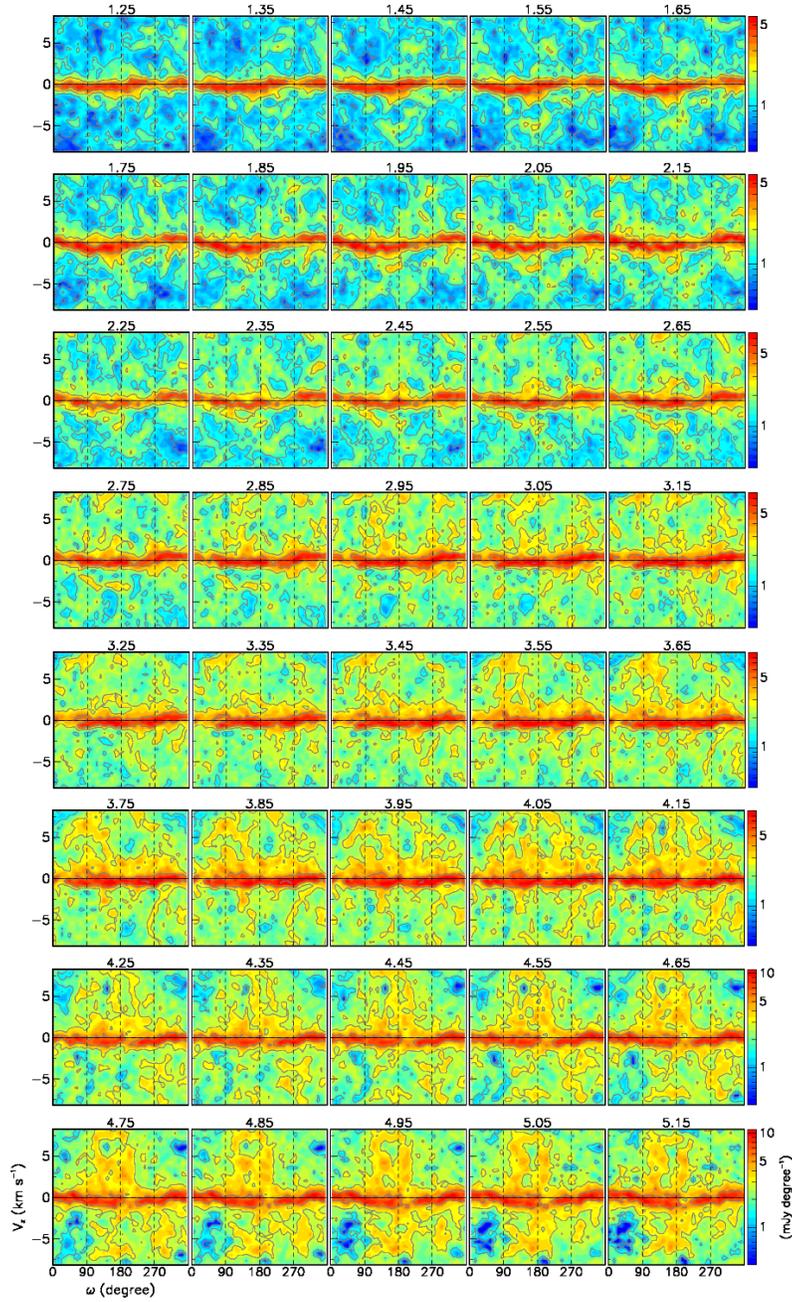

Fig. A4: Projections of the data cube of CO(2-1) emission (combined data) in 40 intervals of $R$ (the central value in arcsec is indicated on top of each panel) on the $V_z$ vs $\omega$ plane ($|V_z| < 8$ km s$^{-1}$, $0 < \omega < 360°$).

**SUPPLEMENTARY MATERIAL**

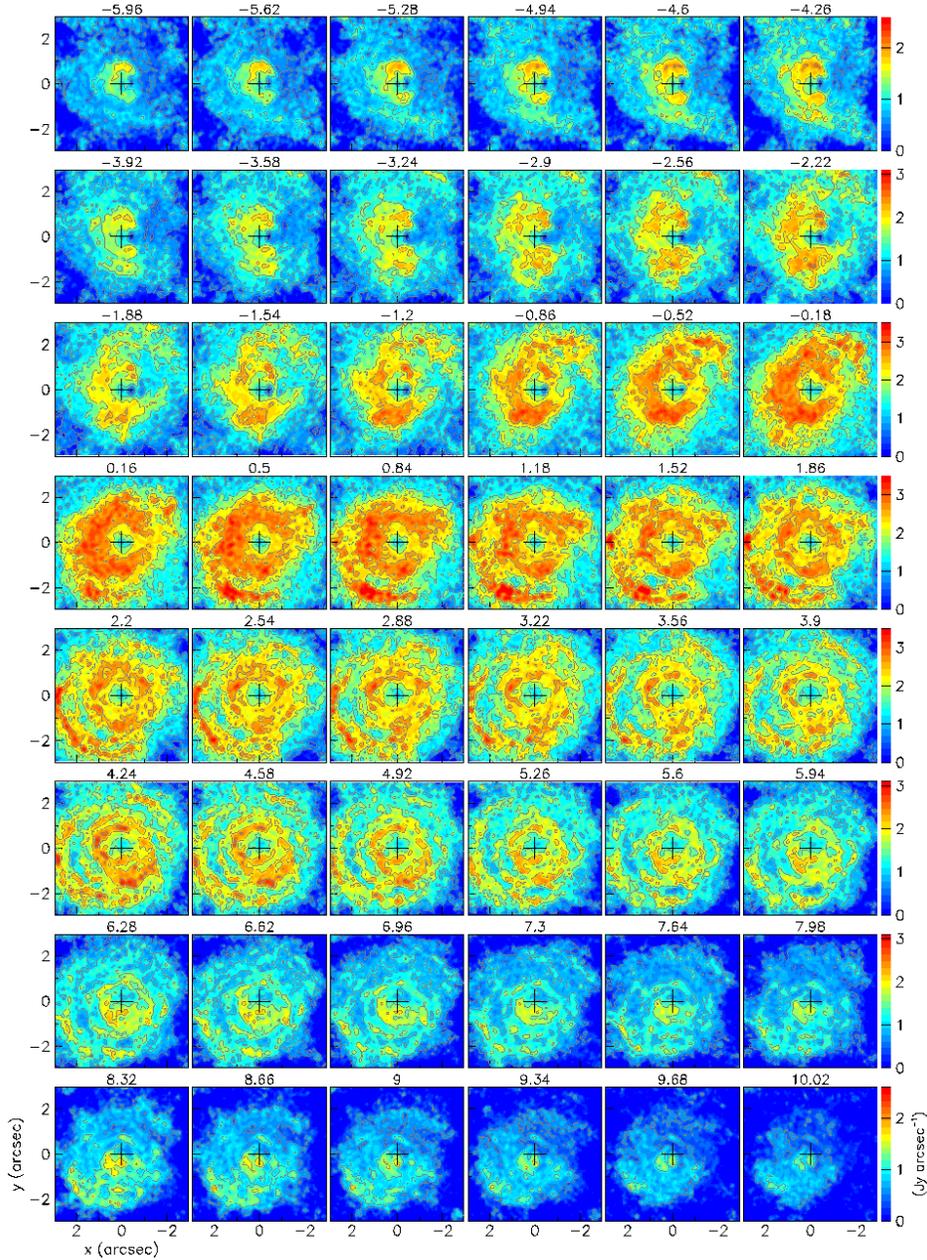

Fig. S1: Channel maps of the SiO(5-4) line emission (combined data). Brightness is multiplied by $R$. Central Doppler velocities (km s$^{-1}$) are indicated for each channel on top of the panel.

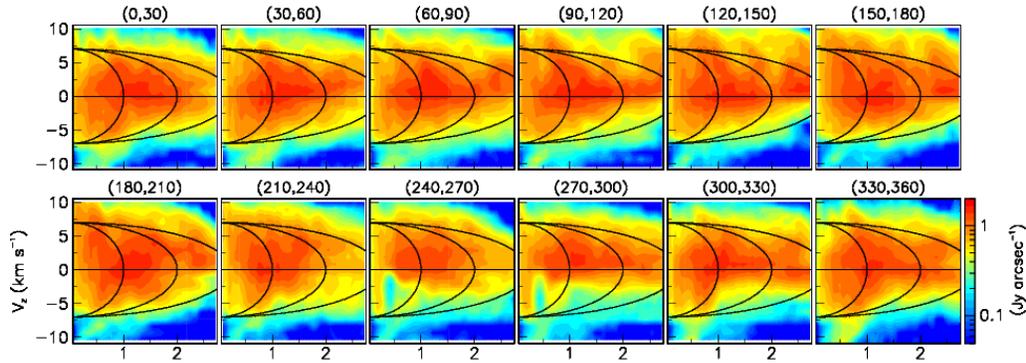

Fig. S2: SiO(5-4) emission (combined data): $V_z$ vs $R$ maps in 12 intervals of position angle $\omega$ as indicated on top of each panel. Ellipses corresponding to spherical shells having radius of 1, 2 and 3 arcsec expanding at uniform radial velocity of 7 km s$^{-1}$ are shown as references.

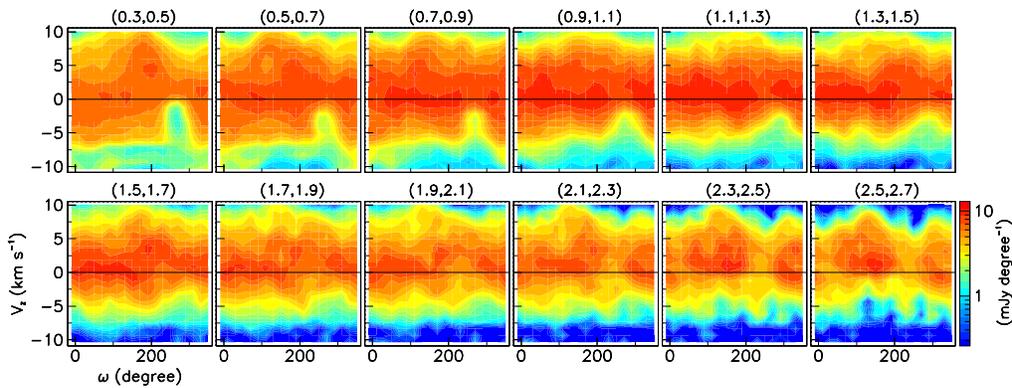

Fig. S3: SiO(5-4) emission (combined data): $V_z$ vs $\omega$ maps in 12 intervals of $R$ (arcsec) as indicated on top of each panel.

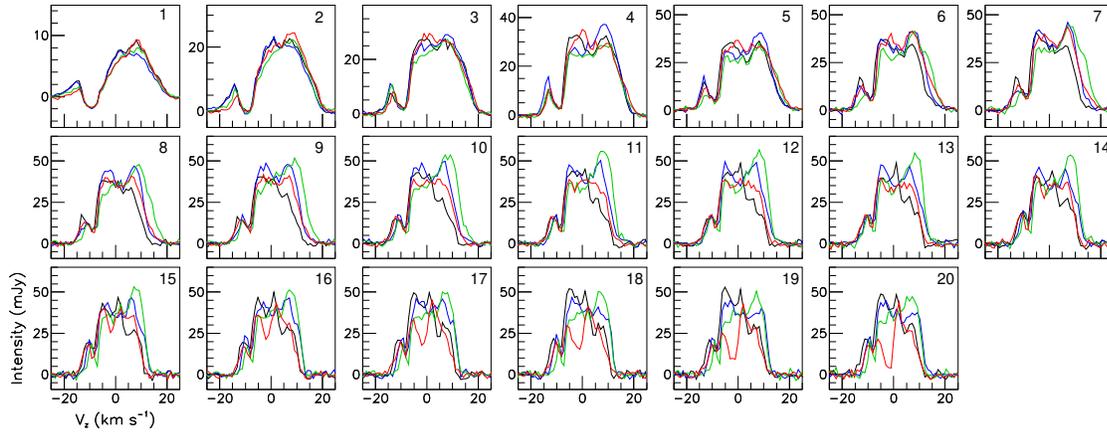

Fig. S4: Doppler velocity spectra of SiO(5-4) line emission (TM0 data). Spectra are integrated in 4 quadrants (north (black), east (blue), south (green) and west (red)) of 20 rings, each 20 mas wide with mean radii increasing from 10 mas (panel 1) to 390 mas (panel 20).

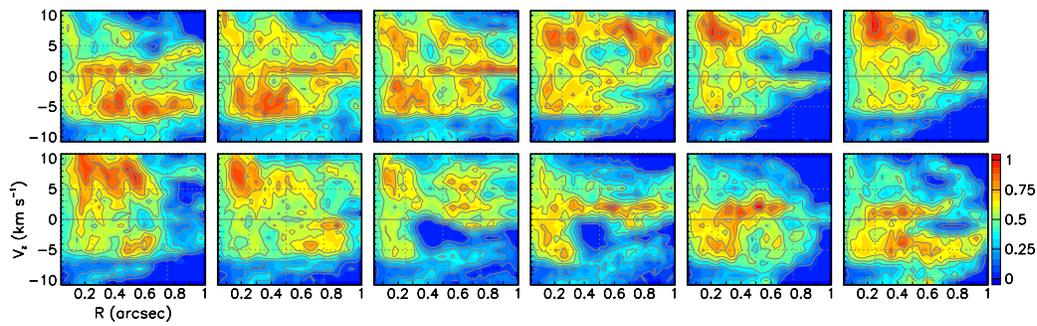

Fig. S5: SiO(5-4) emission, TM0 data. $V_z$ vs $R$ maps in twelve 30° wide sectors of $\omega$ with $\omega$ increasing from upper-left to lower-right panels.

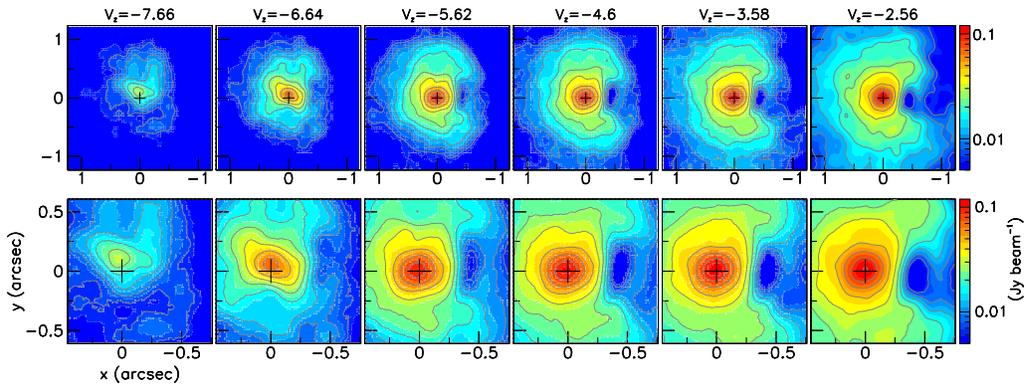

Fig. S6: Channel maps of combined SiO(5-4) data relevant to the region of western depression. The lower row is a zoom of the upper row.

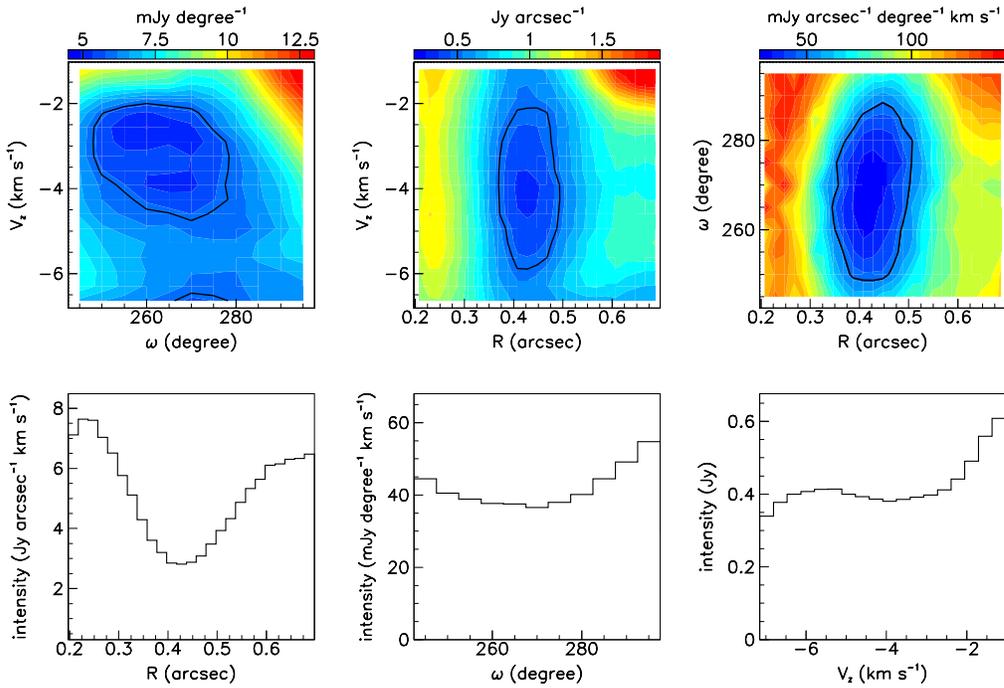

Fig. S7: SiO(5-4) emission, combined data. Projections of the data cube volume defined as $0.2 < R < 0.7$ arcsec, $240° < \omega < 300°$, $-7 < V_z < -1$ km s$^{-1}$.